\documentclass[superscriptaddress,amsmath,amssymb,amsthm,preprint]{revtex4}
\usepackage{amsmath}
\usepackage{graphicx}
\usepackage{epstopdf}
\usepackage{float}
\usepackage{hyperref}
\usepackage{subfigure}
\usepackage{color}
\usepackage[T1]{fontenc}
\usepackage[utf8]{inputenc}
\usepackage[toc,page]{appendix}
\usepackage[usenames,dvipsnames]{xcolor}
\usepackage[normalem]{ulem}
\usepackage{amssymb}
\usepackage{mathrsfs}
\usepackage{grffile}
\usepackage{ulem}

\begin{document}
\title{Topological classes of higher-dimensional black holes in massive gravity}

\author{Deyou Chen}
\email{deyouchen@hotmail.com}
\affiliation{School of Science, Xihua University, Chengdu 610039, China}
\affiliation{Key Laboratory of High Performance Scientific Computation, Xihua University, Chengdu 610039, China}

\author{Yucheng He}
\email{heyucheng365@hotmail.com}
\affiliation{School of Science, Xihua University, Chengdu 610039, China}
\affiliation{Key Laboratory of High Performance Scientific Computation, Xihua University, Chengdu 610039, China}

\author{Jun Tao}
\email{taojun@scu.edu.cn}
\affiliation{Center for Theoretical Physics, College of Physics, Sichuan University, Chengdu, 610065, China}

\

\begin{abstract}
In this paper, we study topological numbers for five-, six- and seven-dimensional anti-de Sitter black holes in the ghost-free massive gravity. We find that when the black holes are charged, they have the same topological number. The topological numbers for the uncharged black holes are 0 or 1, and the specific values are determined by the values of the black holes' parameters. Since $k$ and $c_ 0^2c_2 m^2 $ appear together in the generalized free energy in the form of $k +c_ 0^2c_2 m^2 $, where $k$ characterizes the horizon curvature and $c_2 m^2 $ is the coefficient of the second term of massive potential associated with the graviton mass, this result is applicable to the black holes with the spherical, Ricci flat or hyperbolic horizons. This work shows that the parameters of the ghost-free massive gravity play an important role in topological classes of black holes.
\end{abstract}

\maketitle
\newpage

\section{Introduction}

Topology is an important mathematical tool. When using this tool, the details of the research objects are ignored, and only their generic properties are focused on. The generic properties are revealed by topological quantities. Cunha, Berti and Herdeiro first used the topological approach to study light rings of ultracompact objects (UCOs) \cite{CBH1}. They found that an UCO must have at least two light rings, one of them is stable and the other is unstable. This approach was extended to the four-dimensional, axisymmetric asymptotically flat black hole and a significant result was obtained in \cite{CBH2}. The result showed that there is at least one standard light ring outside the black hole for each rotation sense. The radial stability of light rings is very important for black holes' shadows, and non-extremal and extremal black holes exist in our universe. Given these, some important improvement was implemented and universal properties of light rings(spheres) were gotten in \cite{GG1,SWW1,GG2,GG3}. In \cite{GG4}, the authors also put forward an topological approach to study the black hole shadow caused by the existence of the light rings or light spheres.

Duan's topological current $\phi$-mapping theory is famous for its wide application in various physical systems \cite{DL1,DL2}. Using his theory, Wei, Liu and Mann proposed another important topological approach to research local and global properties of black holes \cite{WLM}. In this approach, the black hole solutions are seen as defects in the thermodynamic parameter space. The defects are related to zero points of a field at $\vec{x} = \vec{z}$ in a space and researched in terms of winding numbers. The winding numbers reveals local thermodynamic characteristics of black holes. The positive/negative numbers characterize the local stability/instability of the black holes. The global properties are determined by the topological charges. Each black hole is endowed with a topological charge, and then these black holes can be divided into different classes according to the values of the charges. A key point in \cite{WLM} is that a vector field was constructed with a generalized free energy. A similar generalized free energy was defined in \cite{LW1,JWY1,JWY2} and used to research on the black hole phase transitions. Since this approach was proposed, it has immediately attracted people's attention. Based on this approach, some new viewpoints on the topological approach have been proposed, the topological arguments on the black holes in the complex spacetimes or in the different ensembles were performed \cite{SWW2,SWW3,YBM,BLT,CWL1,ZYF,FJW,DWW1,DWW2,DWW3,DWW4,LLZW,DZ1,DZ2,DZ3,LW2,GP1,GP2,GP3, ZCHLY,AAGS,XWWY,HMM}. These work strongly supports the viewpoint proposed in \cite{WLM} that the topological number for the black hole branches at an arbitrary given temperature is a universal number independent of the black hole's parameters. These researches are shining with new light on black holes physics.

In this paper, we study the topological numbers for high-dimensional, charged and uncharged black holes in the ghost-free massive gravity, specifically five-, six- and seven-dimensional black holes. In the calculation, we change the values of the black holes' parameters and calculate their topological numbers. We find that the numbers for the charged black holes are same, while the numbers for the uncharged black holes depend on the values of their parameters. Einstein's general relativity (GR) is a low energy effective theory. The UV completeness requires that GR be modified to meet physical descriptions in the high energy region. Massive gravity is one of the modified gravity theories. In this gravity, the graviton is endowed with mass. A spherically symmetric black hole solution with a negative cosmological constant was obtained in \cite{DV}, and its thermodynamical properties in the extended phase space were researched in \cite{CHPZ,XCH1,XCH2,DYC}.

The rest of this paper is organized as follows. In the next section, we briefly review the solution of the higher-dimensional black holes in the massive gravity and discuss its thermodynamical properties. In Sec. \ref{Sec3}, we calculate the topological numbers for the charged and uncharged black holes by changing the values of the black holes' parameters. Sec. \ref{Sec4} is devoted to our conclusion and discussion.

\section{Higher-dimensional black hole solution in massive gravity}\label{Sec2}

The action for an $(n + 2)$-dimensional ghost-free massive gravity is given by \cite{DV}

\begin{eqnarray}
\mathcal{S} &=& \frac{1}{16\pi}\int{dx^{n+2}\sqrt{-g}\left[R +\frac{n(n+1)}{l^2}-\frac{F^2}{4}+m^2\sum_{i=1}^4{c_iu_i(g,f)}\right]},
\label{eq2.1}
\end{eqnarray}

\noindent where $F = F^{\mu\nu}F_{\mu\nu}$ and  $F_{\mu\nu}$ is the electromagnetic tensor, the terms including $m^2$ are the massive potential associate with graviton mass, $f$ is a fixed symmetric tensor called as the reference metric. $c_i$ are constants, and their values are discussed in \cite{CHPZ}. $u_i$ are symmetric polynomials of the eigenvalues of the $(n+2)\times (n+2)$ matrix $\mathcal{K}_{\nu}^{\mu}=\sqrt{f^{\mu \alpha}g_{\alpha \nu}}$: $u_1 = [\mathcal{K}]$, $u_2=[\mathcal{K}]^2-[\mathcal{K}^2]$, $ u_3 = [\mathcal{K}]^3-3[\mathcal{K}][\mathcal{K}^2]+2[\mathcal{K}^3]$ ,
$u_4 = [\mathcal{K}]^4-6[\mathcal{K}^2][\mathcal{K}]^2+8[\mathcal{K}^3][\mathcal{K}]+3[\mathcal{K}^2]^2-6[\mathcal{K}^4]$. The square root in $\mathcal{K}$ denotes $(\sqrt{A})_{\nu}^{\mu}(\sqrt{A})_{\lambda}^{\nu}=A_{\lambda}^{\mu}$ and $[\mathcal{K}]= \mathcal{K}_{\mu}^{\mu}$. This action admits a solution of static black hole with the spacetime metric \cite{CHPZ}

\begin{eqnarray}
ds^2 = -f(r)dt^2 + \frac{1}{f(r)}dr^2 + r^2 h_{ij}dx^idx^j,
\label{eq2.2}
\end{eqnarray}

\noindent and reference metric $f_{\mu\nu}=diag(0,0,c_0^2h_{ij})$, where $c_0$ is a positive constant, and $h_{ij}dx^idx^j$ is the line element for an Einstein space with the constant curvature $n(n-1)k$. $k = 1$, $0$ or $-1$ corresponds to a spherical, Ricci flat or hyperbolic topology horizons of the black hole, respectively. According to the reference metric, we get the symmetric polynomials

\begin{eqnarray}
u_1 &=& \frac{nc_0}{r}, \quad \quad u_2=\frac{n(n-1)c_0^2}{r^2},
\nonumber\\
u_3 &=& \frac{n(n-1)(n-2)c_0^3}{r^3}, \quad u_4 = \frac{n(n-1)(n-2)(n-3)c_0^4}{r^4}.
\label{eq2.3}
\end{eqnarray}

\noindent The expression of the metric function $f(r)$ is

\begin{eqnarray}
f(r)&=& k+\frac{r^2}{l^2}-\frac{16\pi M}{nV_nr^{n-1}}+\frac{(16\pi Q)^2}{2n(n-1)V_n^2r^{2(n-1)}}+\frac{c_0c_1m^2r}{n}+c_0^2c_2m^2\nonumber\\
&&+\frac{(n-1)c_0^3c_3m^2}{r} +\frac{(n-1)(n-2)c_0^4c_4m^2}{r^2},
\label{eq2.4}
\end{eqnarray}

\noindent where $l^2$ is related to the cosmological constant $\Lambda$ as $l^2 = -\frac{n(n+1)}{2\Lambda }$. $M$ and $Q$ are the black hole's mass and charge, respectively. $V_n$ is the volume spanned by coordinates $x^i$. This black hole describes a static charged spacetime. The event horizon  $r_h$ is determined by $f (r)=0$. We rewrite the mass with the horizon and charge as follows

\begin{eqnarray}
M &=& \frac{nV_n r_h^{n-1}}{16\pi} \left[k+\frac{r_h^2}{l^2}+\frac{(16\pi Q)^2}{2n(n-1)V_n^2r_h^{2(n-1)}}+\frac{c_0c_1m^2r_h}{n}+c_0^2c_2m^2 \right. \nonumber\\
&&\left.+\frac{(n-1)c_0^3c_3m^2}{r_h} +\frac{(n-1)(n-2)c_0^4c_4m^2}{r_h^2}\right].
\label{eq2.5}
\end{eqnarray}

The entropy, electric potential and Hawking temperature at the event horizons are given by

\begin{eqnarray}
S &=&\frac{V_nr_h^{n}}{4}, \quad \quad\quad  \Phi_e= \frac{16\pi Q}{(n-1)V_nr_h^{n-1}},\nonumber\\
T &=& \frac{1}{4\pi r_+} \left[\frac{(n+1)r_h^2}{l^2}+\frac{(16\pi Q)^2}{2nV_n^2r_h^{2(n-1)}} + c_0c_1m^2r_h + (n-1)c_0^2c_2m^2 +(n-1)k\right. \nonumber\\
&&\left.+\frac{(n-1)(n-2)c_0^3c_3m^2}{r_h} +\frac{(n-1)(n-2)(n-3)c_0^4c_4m^2}{r_h^2}\right],
\label{eq2.6}
\end{eqnarray}

\noindent respectively. The thermodynamics in the extended phase space of the massive gravity were studied in \cite{CHPZ,XCH1,XCH2,DYC}, where the cosmological constant was seen as a variable related to pressure, $P=-\frac{\Lambda}{8\pi} =\frac{n(n+1)}{16\pi l^2}$, and its conjugate quantity is a thermodynamic volume $V$. Then the mass is regarded as the enthalpy \cite{KRT1,KRT2,KRT3,KRT4,KRT5}. For this black hole, its thermodynamic volume is $ V= \frac{V_nr_h^{n+1}}{n+1}$.  It was found that the van der Waals-like phase transition exists the black hole not only with the spherical black holes but also with the Ricci flat and hyperbolic horizons. When $c_1$, $c_2$, $c_3$ and $c_4$ are seen as extensive parameters for the mass, and their conjugate quantities are

\begin{eqnarray}
\Phi_1 &=& \frac{V_nc_0m^2r_h^{n}}{16\pi},\quad \quad\quad \quad\quad\quad \Phi_2 = \frac{nV_nc_0^2m^2r_h^{n-1}}{16\pi},\nonumber\\
\Phi_3 &=& \frac{n(n-1)V_nc_0^3 m^2r_h^{n-2}}{16\pi}, \quad\quad  \Phi_4 = \frac{n(n-1)(n-2)V_nc_0^4 m^2r_h^{n-3}}{16\pi}.
\label{eq2.7}
\end{eqnarray}

\noindent Obviously, these quantities obey the first law of thermodynamics

\begin{eqnarray}
dM=TdS+VdP+\Phi_edQ+ \sum_{i=1}^4 \Phi_idc_i.
\label{eq2.8}
\end{eqnarray}

\noindent When the cosmological constant and parameters $c_i$ are fixed, the terms $VdP$ and $\Phi_idc_i$ disappear and the mass is the internal energy.

\section{Topological properties of higher-dimensional black holes in massive gravity}\label{Sec3}

In this section, we first review the topological approach proposed in \cite{WLM}, and then change the values of the black holes' parameters to study the topological numbers for the different higher-dimensional black holes in the ghost-free massive gravity. 

\subsection{Topological current}
According to Ref. \cite{WLM}, the generalized free energy is defined by

\begin{eqnarray}
\mathcal{F} = E- \frac{S}{\tau},
\label{eq3.1}
\end{eqnarray}

\noindent where $E$ and $S$ are the energy and entropy of a black hole system, respectively. $\tau$ is a variable and can be seen as the inverse temperature of the cavity enclosing the black hole. This free energy is off-shell except at $\tau = \frac{1}{T}$. To calculate the topological number, a vector is constructed as follows

\begin{eqnarray}
\phi = \left(\frac{\partial \mathcal{F}}{\partial r_h} , -\cot\Theta \csc\Theta\right).
\label{eq3.2}
\end{eqnarray}

\noindent In this vector, $0<r_h<+\infty$ and $0\le\Theta\le\pi$. Zero points of the vector obtained at $\tau = 1/T$ and $\Theta = \pi/2$ correspond to the on-shell black hole solution. Other points are not the solutions of Einstein field equations, and they are the off-shell states. $\phi^{\Theta}$ diverges at $\Theta =0$ and $\Theta =\pi$, which leads to that the direction of the vector is outward.

Using Duan's $\phi$-mapping topological current theory \cite{DL1,DL2}, we define a topological current

\begin{eqnarray}
j^{\mu} = \frac{1}{2\pi}\varepsilon^{\mu\nu\rho}\varepsilon_{ab}\partial_{\nu}n^a\partial_{\rho}n^b,
\label{eq3.3}
\end{eqnarray}

\noindent where $\mu,\nu,\rho = 0,1,2$, $a,b = 1,2$, $\partial_{\nu} = \frac{\partial}{x^{\nu}}$ and $x^{\nu} = (\tau, r_h, \Theta)$. $\tau$ is seen as a time parameter of the topological defect. $n^a$ is a unit vector defined by $\left( \frac{n^r}{||n||}, \frac{n^{\Theta}}{||n||}\right)$. It is easily to prove the current is conserved. Using the Jacobi tensor $\varepsilon^{ab}J^{\mu}(\frac{\phi}{x}) = \varepsilon^{\mu\nu\rho} \partial_{\nu}{\phi}^a\partial_{\rho}{\phi}^b$ and two-dimensional Laplacian Green function $\Delta_{\phi^a}ln||\phi||=2\pi \delta^2(\phi)$, the current is rewritten as

\begin{eqnarray}
j^{\mu} = \delta^2(\phi)j^{\mu}\left(\frac{\phi}{x}\right),
\label{eq3.4}
\end{eqnarray}

\noindent which is nonzero only when $\phi^a(x^i) = 0$. Then the number in a parameter region $\sum$ is calculated as

\begin{eqnarray}
W= \int_{\sum}j^{0}d^2x= \sum_{i=1}^{N}\beta_i\eta_i = \sum_{i=1}^{N}w_i,
\label{eq3.5}
\end{eqnarray}

\noindent where $j^{0} = \sum_{i=1}^{N}\beta_i\eta_i\delta^2(\vec{x}-\vec{z}_i)$ is the density of the current. $\beta_i$ is Hopf index which counts the number of the loops that $\phi^a$ makes in the vector $\phi$ space when $x^{\mu}$ goes around the zero point $z_i$. Clearly, this index is always positive. $\eta_i$=sign$J^0(\phi/x)_{z_i} = \pm 1$ is the Brouwer degree. $w_i$ are the winding number for the $i$-th zero point of the vector in the region and their values are independent on the shape of the region. In recent researches, people found these values are determined by the (un-)stable black holes. 

\subsection{Topological numbers for five-dimensional black holes}

When $n=3$, the metric (\ref{eq2.2}) describes a five-dimensional charged spacetime. According to Eqs. (\ref{eq2.5}) and (\ref{eq2.6}), the mass and entropy are

\begin{eqnarray}
M=\frac{3 V_3 r_h^2}{16 \pi }\left[ k+ \frac{c_0 c_1 m^2 r_h}{3}+c_2 c_0^2 m^2+\frac{2 c_3 c_0^3 m^2}{r_h}+\frac{4\pi Pr_h^2}{3}+\frac{(16 \pi  Q)^2}{12 V_3^2 r_h^4}+\frac{6c_0^4c_4m^2}{r_h^2}\right],
\label{eq3.1.1}
\end{eqnarray}

\begin{eqnarray}
 S=\frac{1}{4} V_3 r_h^3,
\label{eq3.1.2}
\end{eqnarray}

\noindent respectively. Since the cosmological constant is a fixed constant, the black hole's energy is its mass. Inserting Eqs. (\ref{eq3.1.1}) and (\ref{eq3.1.2}) into Eq.(\ref{eq3.1}), we get the off-shell free energy

\begin{eqnarray}
\mathcal{F} &=&\frac{3 V_3 r_h^2}{16 \pi }\left[ k+ \frac{c_0 c_1 m^2 r_h}{3}+c_2 c_0^2 m^2+\frac{2 c_3 c_0^3 m^2}{r_h}+\frac{4\pi Pr_h^2}{3}+\frac{(16 \pi  Q)^2}{12 V_3^2 r_h^4}+\frac{6c_0^4c_4m^2}{r_h^2}\right] \nonumber\\
&&-\frac{V_3 r_h^3}{4 \tau }.
\label{eq3.1.3}
\end{eqnarray}

\noindent Then the components of the vector $\phi$ are calculated as follows

\begin{eqnarray}
\phi ^{r_h}&=&\frac{3 V_3 r_h \left(c_2 c_0^2 m^2+k\right)}{8 \pi }+\frac{3 c_1 c_0 m^2 V_3 r_h^2}{16 \pi }+\frac{3 c_3 c_0^3 m^2 V_3}{8 \pi }+2 P V_3 r_h^3-\frac{8 \pi  Q^2}{V_3 r_h^3}-\frac{3 V_3 r_h^2}{4 \tau }, \nonumber\\
\phi ^{\Theta} &=& -\cot\Theta \csc\Theta.
\label{eq3.1.4}
\end{eqnarray}

\noindent Zero points are determined by $\phi^{r_h} = 0$. We solve it and get

\begin{eqnarray}
\tau = \frac{12 \pi  V_3^2 r_h^5}{32 \pi  P V_3^2 r_h^6 + 3 c_0 c_1 m^2 V_3^2 r_h^5+6 (c_0^2 c_2 m^2+k) V_3^2 r_h^4+6 c_0^3 c_3 m^2 V_3^2 r_h^3 -128 \pi ^2 Q^2}.
\label{eq3.1.5}
\end{eqnarray}

\begin{figure}[h]
	\centering
	\begin{minipage}[t]{0.48\textwidth}
		\centering
		\includegraphics[scale=0.3]{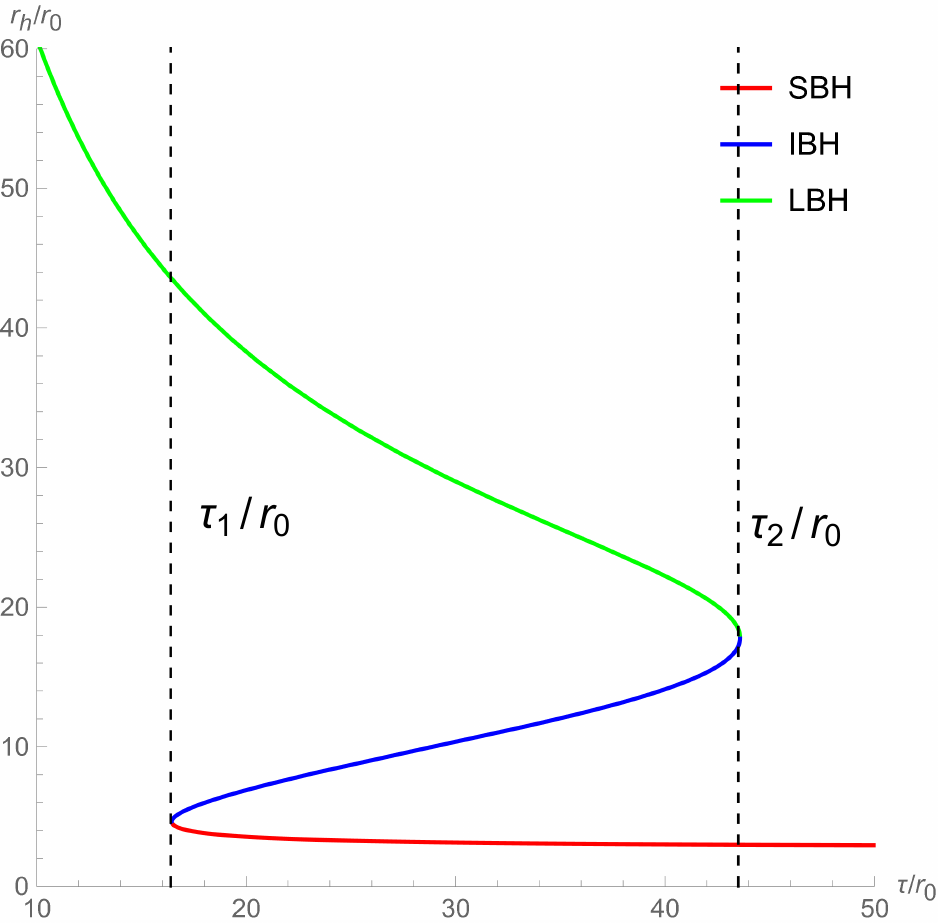}
	\end{minipage}
	\begin{minipage}[t]{0.48\textwidth}
		\centering
		\includegraphics[scale=0.3]{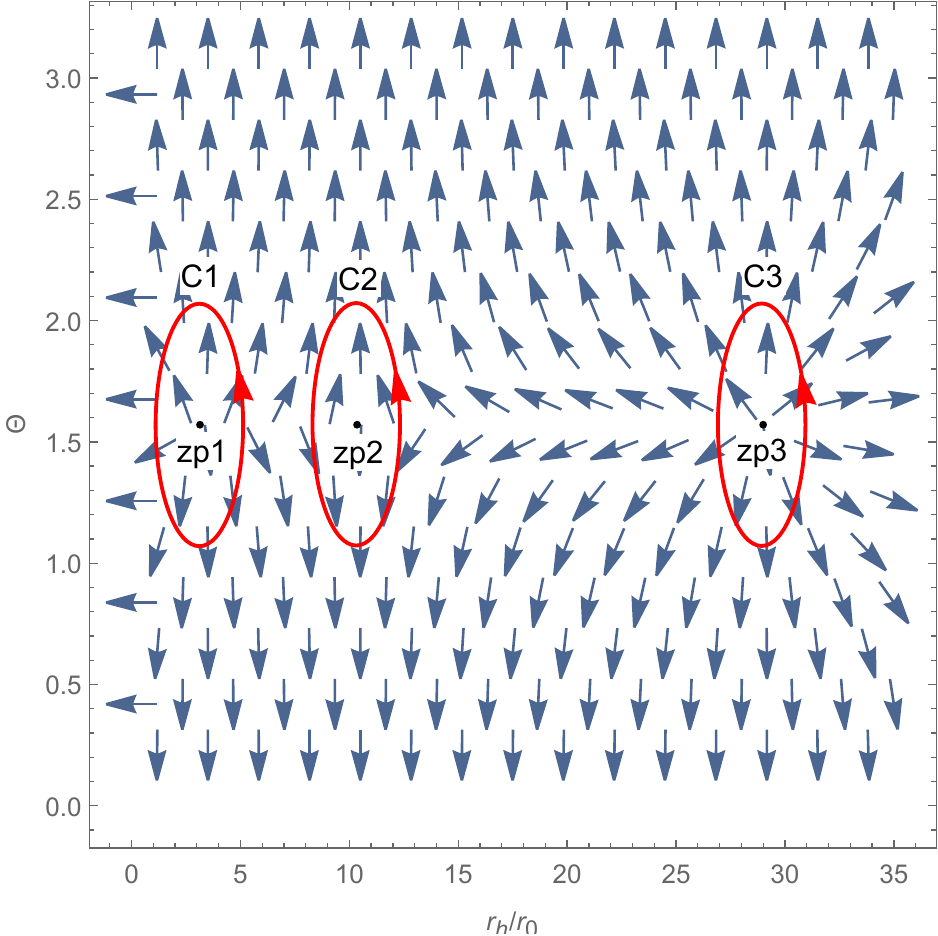}
	\end{minipage}
\caption{Topological properties of the five-dimensional charged black hole, where $a_5 =1/(128\pi ^2)$, $ b_5 = -1$,  $c_5 = 6$, $ k=1$, $d_5 =-15$ and $ Pr_0^{2}=0.001$. Zero points of the vector $\phi^{r_h}$ in the plane $r_h - \tau$ are plotted in the left picture. The unit vector field $n$ on a portion of the plane $r_h - \Theta$ at $\tau /r_0=30 $ is plotted in the right picture. Zero points are at $(r_h/r_0 ,\Theta)$= ($3.15, \pi/2$), ($10.36, \pi/2$) and ($28.99, \pi/2$), respectively.}	
	\label{fig:3.1}
\end{figure}

\begin{figure}[h]
	\centering
	\begin{minipage}[t]{0.48\textwidth}
		\centering
		\includegraphics[scale=0.3]{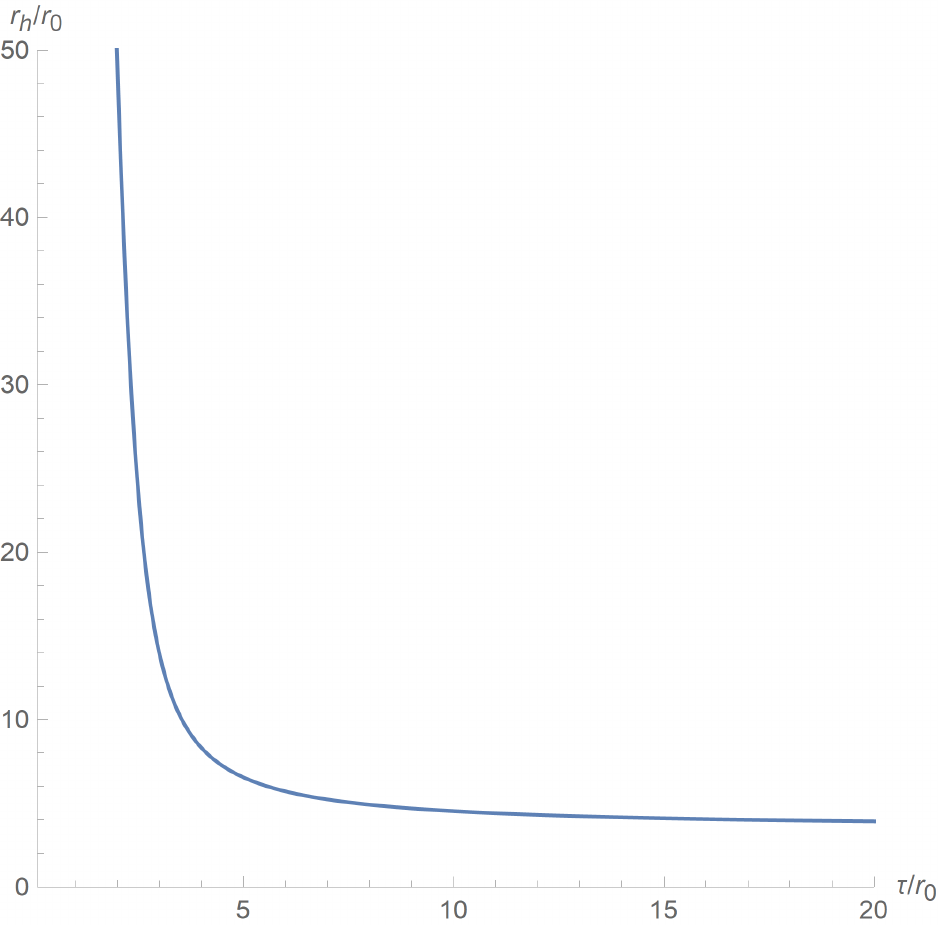}
	\end{minipage}
	\begin{minipage}[t]{0.48\textwidth}
		\centering
		\includegraphics[scale=0.3]{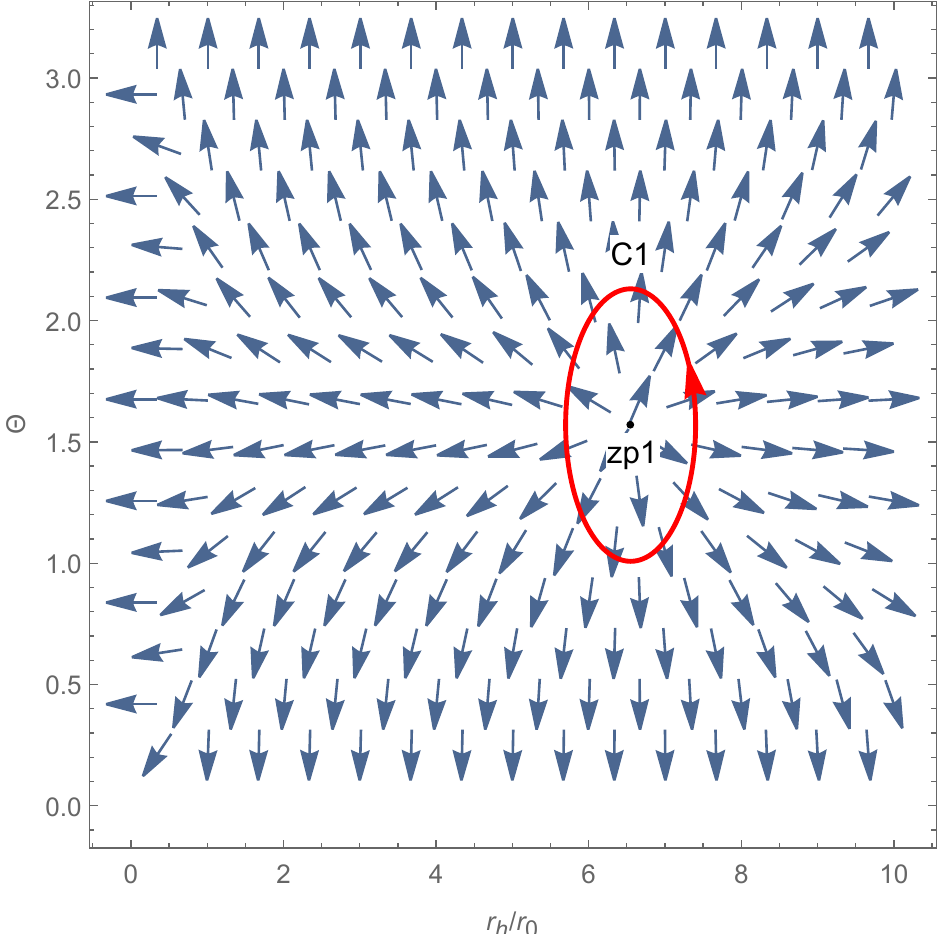}
	\end{minipage}
\caption{Topological properties of the five-dimensional charged black hole, where $a_5 =\frac{1}{128\pi ^2}$, $b_5  = 5$, $c_5 =-10$, $k=1$, $d_5 =1$ and $Pr_0^{2}=0.001$. Zero points of the vector $\phi^{r_h}$ in the plane $r_h - \tau$ are plotted in the left picture. The unit vector field $n$ on a portion of the plane $r_h - \Theta$ at $\tau /r_0=5 $ are plotted in the right picture. The zero point is at $(r_h/r_0 ,\Theta)$= ($6.55, \pi/2$).}	
	\label{fig:3.2}
\end{figure}

\noindent For convenience, we order $a_5 = Q^2/(V_3^2r_0^{4})$, $ b_5 = c_0 c_1 m^2 r_0 $,  $c_5 = c_0^2 c_2 m^2$, $d_5 = c_0^3 c_3 m^2/r_0$. To show the zero points in the plane $r_h - \tau$, we fix the black hole's parameters at certain values. In the left picture of Figure \ref{fig:3.1}, there are three black hole branches when $\tau_1<\tau<\tau_2$, a large black hole (LBH) branch when $\tau<\tau_1$, and a small black hole (SBH) branch when $\tau>\tau_2$. A SBH/LBH phase transition occurs in this region $\tau_1<\tau<\tau_2$. For the LBH region and the SBH region, the winding numbers for the zero points are $w=+1$. For the intermediate black hole (IBH) region, the winding number for the zero points is $w=-1$. It has been proved that the positive/negative winding numbers imply the thermodynamic stability/instability. Thus the number of this black hole is $W=1$. We order $\tau/r_0=30$ and plot the unit vector field $n$ in the right picture of Figure 1. There are three zero points, zp1, zp2 and zp3 in the picture, and they are at $(r_h/r_0 ,\Theta)$= ($3.15, \pi/2$), ($10.36, \pi/2$) and ($28.99, \pi/2$), respectively. Their winding numbers are 1, -1 and 1, respectively. Then the topological number is also $1$.

To further study the topological properties of this black hole, we draw Figure \ref{fig:3.2}. Obviously, $r_h$ monotonically decreases with the increase of the inverse temperature $\tau$, which implies that there is only a stable black hole and no phase transition occurs. There is only one zero point in the right picture of Figure \ref{fig:3.2}. Then the topological number is $1$, which is completely consistent with the result obtained in Figure \ref{fig:3.1}. Although the black hole branches and numbers of zero points in Figure \ref{fig:3.1} and Figure \ref{fig:3.2} are different, these data all indicate the same topological number.

\begin{figure}[h]
	\centering
	\begin{minipage}[t]{0.48\textwidth}
		\centering
		\includegraphics[scale=0.3]{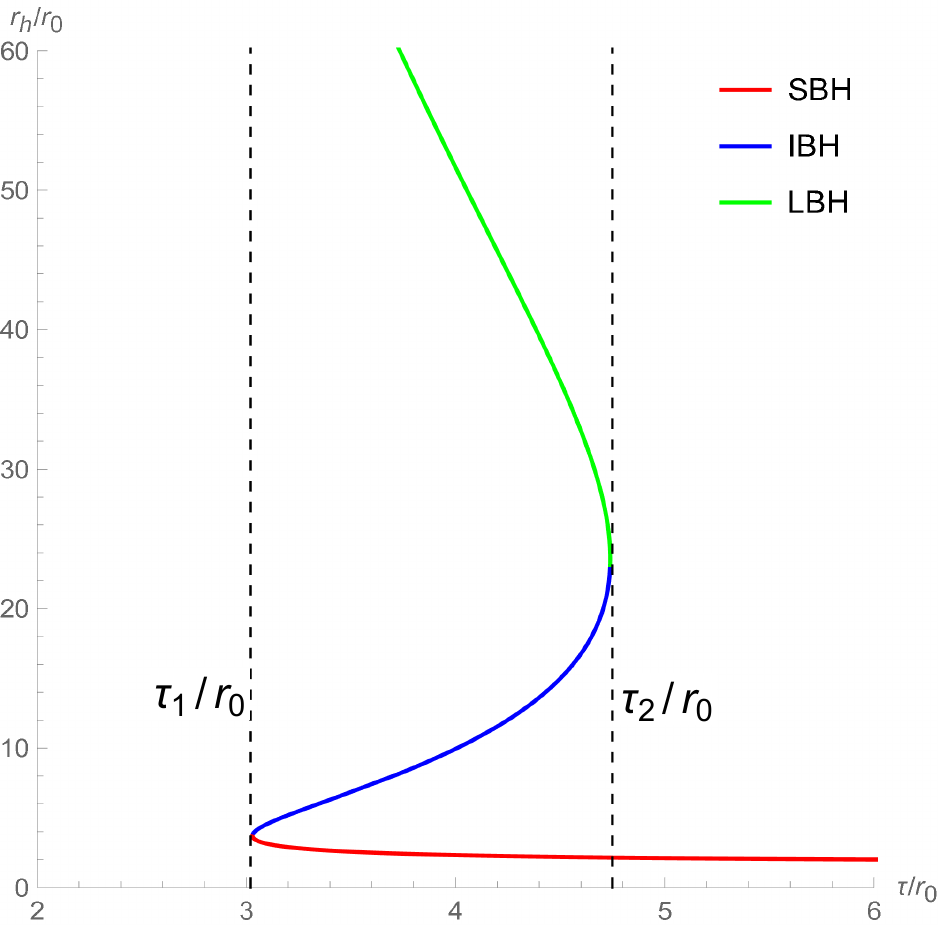}
	\end{minipage}
	\begin{minipage}[t]{0.48\textwidth}
		\centering
		\includegraphics[scale=0.3]{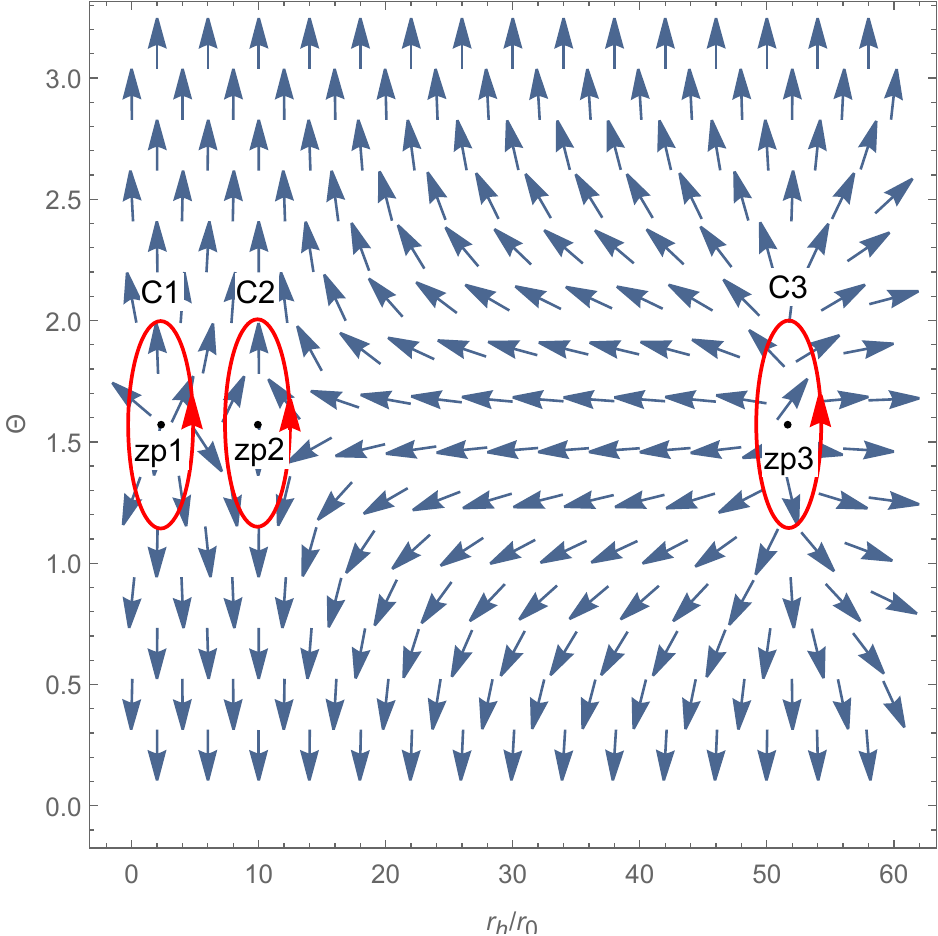}
	\end{minipage}
\caption{Topological properties of the five-dimensional uncharged black hole, where $a_5=0$, $b_5 = 1$, $c_5=-10$, $k=1$, $d_5=-20$ and $Pr_0^{2}=0.001$. Zero points of the vector $\phi^{r_h}$ in the plane $r_h - \tau$ are plotted in the left picture. The unit vector field $n$ on a portion of the plane $r_h - \Theta$ at $\tau /r_0=4 $ is plotted in the right picture. Zero points are at $(r_h/r_0 ,\Theta)$=($2.33, \pi/2$), ($9.94, \pi/2$) and ($51.64, \pi/2$), respectively.}	
	\label{fig:3.3}
\end{figure}

\begin{figure}[h]
	\centering
	\begin{minipage}[t]{0.48\textwidth}
		\centering
		\includegraphics[scale=0.3]{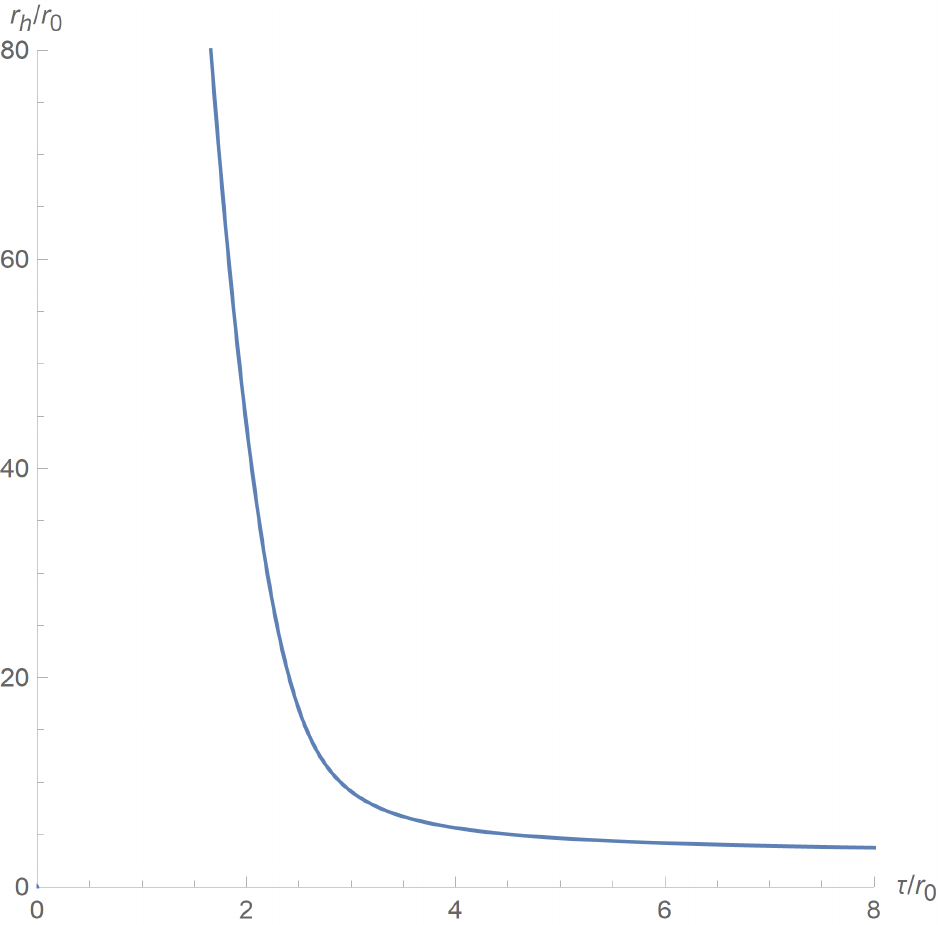}
	\end{minipage}
	\begin{minipage}[t]{0.48\textwidth}
		\centering
		\includegraphics[scale=0.3]{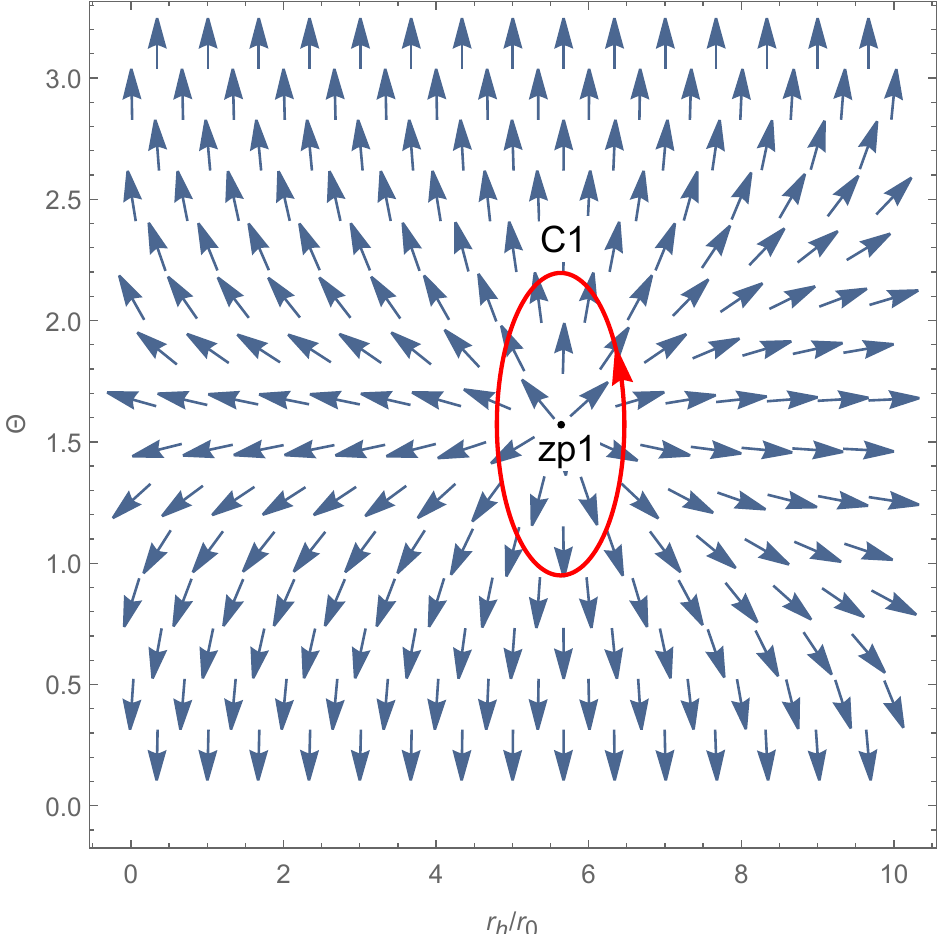}
	\end{minipage}
\caption{Topological properties of the five-dimensional uncharged black hole, where $a_5=0$, $b_5 = 5$, $c_5= -5$, $k=1$, $d_5= -10$ and $Pr_0^{2}=0.001$. Zero points of the vector $\phi^{r_h}$ in the plane $r_h - \tau$ are plotted in the left picture. The unit vector field $n$ on a portion of the plane $r_h - \Theta$ at $\tau /r_0=4 $ is plotted in the right picture. The zero points is at $(r_h/r_0 ,\Theta)$=($5.64, \pi/2$).}	
	\label{fig:3.4}
\end{figure}

\begin{figure}[h]
	\centering
	\begin{minipage}[t]{0.48\textwidth}
		\centering
		\includegraphics[scale=0.3]{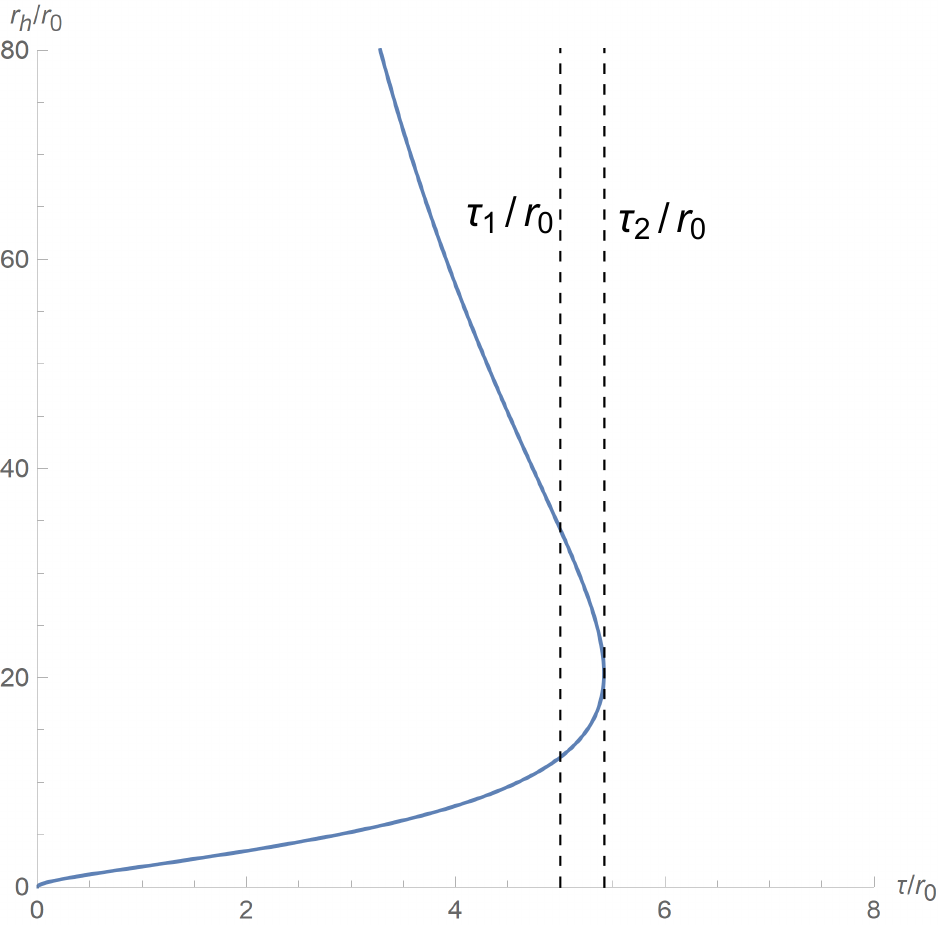}
	\end{minipage}
	\begin{minipage}[t]{0.48\textwidth}
		\centering
		\includegraphics[scale=0.3]{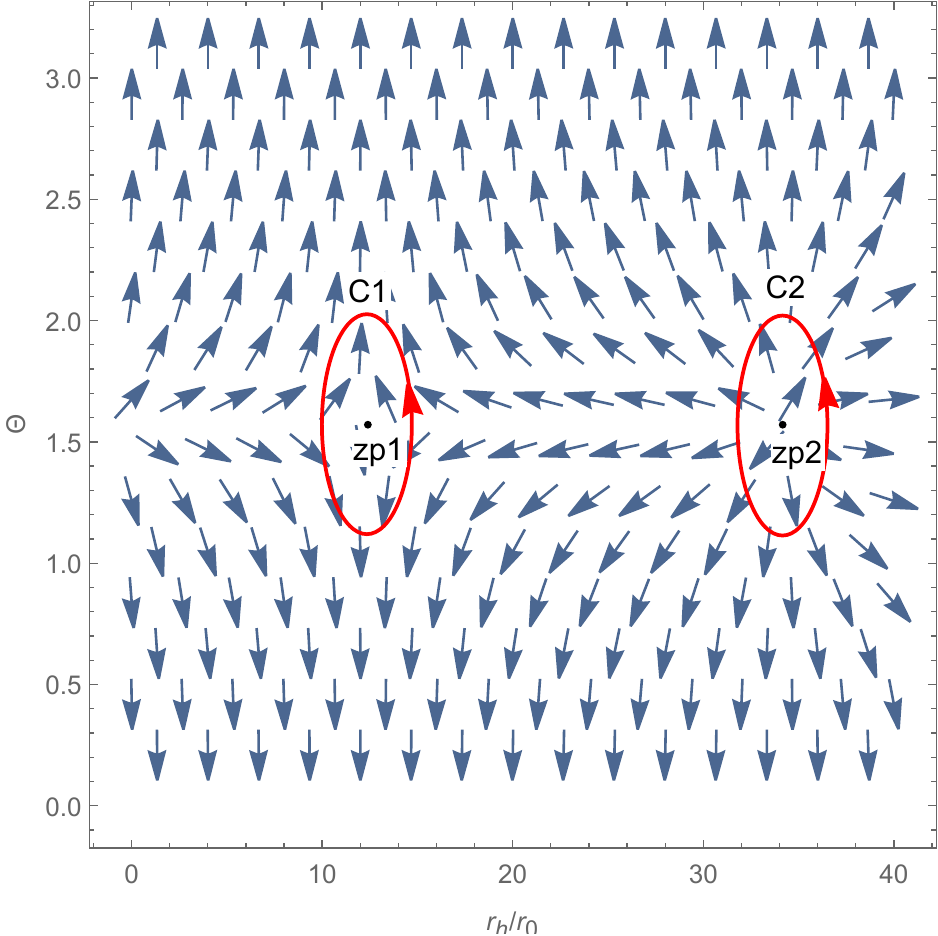}
	\end{minipage}
\caption{Topological properties of the five-dimensional uncharged black hole, where $a_5=0$, $b_5 = 1$, $c_5= 5$, $k=1$, $d_5= 10$ and $Pr_0^{2}=0.001$. Zero points of the vector $\phi^{r_h}$ in the plane $r_h - \tau$ are plotted in the left picture. The unit vector field $n$ on a portion of the plane $r_h - \Theta$ at $\tau /r_0=5 $ is plotted in the right picture. Zero points are at $(r_h/r_0 ,\Theta)$=($12.40, \pi/2$) and ($34.17, \pi/2$), respectively.}	
	\label{fig:3.5}
\end{figure}

Now we order $Q=0$ in Eq. (\ref{eq2.4}) and discuss the topological properties of the uncharged black hole in Figure \ref{fig:3.3} - Figure \ref{fig:3.5}. In the left picture of Figure \ref{fig:3.3}, there are three different black hole branches when $\tau_1<\tau<\tau_2$, which represent the LBH, IBH and SBH. The generation point and annihilation point are located at $\tau_1/r_0$ and $\tau_2/r_0$, respectively. Therefore, there is a SBH/LBH phase transition. For the LBH and SBH branches, their winding numbers for any zero point is $w=1$. For the IBH branch, the number is $w=-1$. Thus the topological number is $1$. There is only a stable black hole and a zero point in Figure \ref{fig:3.4}, which implies there is no phase transition for this black hole in this case. Its topological number is also obtained as $1$, which is full in consistence with the results obtained for the charged black hole.

In Figure \ref{fig:3.5}, the annihilation point is at $\tau_2/r_0$. There are two different black holes at the same temperature when $\tau <\tau_2$. The first black hole with large radius is stable and has the winding number for any zero point is $w=1$, the second black hole with small radius is unstable and has the winding number for any zero point is $w=-1$. The topological number is gotten as $0$, which is different from the result derived in Figure \ref{fig:3.3} and Figure \ref{fig:3.4}. Comparing Figure \ref{fig:3.3}, \ref{fig:3.4} and \ref{fig:3.5}, we find that the topological number for this uncharged black hole is $0 $ or $1$, whose specific values are determined by the values of the black hole's parameters. Therefore, the topological number for the five-dimensional charged black hole is $1$, while that for the uncharged black hole  is $0 $ or $1$.

\subsection{Topological numbers for six-dimensional black holes}

For the six-dimensional black hole, $n=4$ in the metric (\ref{eq2.2}). The mass and entropy are

\begin{eqnarray}
	M=\frac{V_4 r_h^3}{4 \pi }\left[\frac{6 c_4 c_0^4 m^2}{r_h^2}+\frac{3 c_3 c_0^3 m^2}{r_h}+\frac{ c_0 c_1 m^2r_h}{4} +c_2 c_0^2 m^2+\frac{4 \pi  P}{5} r_h^2+\frac{(16 \pi  Q)^2}{12 V_4^2 r_h^6}+k\right],
\label{eq3.2.1}
\end{eqnarray}

\begin{eqnarray}
	S=\frac{1}{4} V_4 r_h^4,
\end{eqnarray}

\noindent respectively. Inserting Eqs. (\ref{eq3.2.1}) and (\ref{eq3.2.2}) into Eq.(\ref{eq3.1}), we get the generalized free energy energy

\begin{eqnarray}
\mathcal{F}=\frac{V_4 r_h^3}{4 \pi }\left[\frac{6 c_4 c_0^4 m^2}{r_h^2}+\frac{3 c_3 c_0^3 m^2}{r_h}+\frac{c_0 c_1 m^2r_h}{4}+c_2 c_0^2 m^2+k+\frac{4 \pi  Pr_h^2}{5}+\frac{(16 \pi  Q)^2}{12 V_4^2 r_h^6}\right]-\frac{V_4 r_h^4}{4 \tau }.
\label{eq3.2.2}
\end{eqnarray}

\noindent We use the definition of the vector and obtain its component 

\begin{eqnarray}
\phi ^{\text{rh}} = \frac{V_4 \left[3 r_h^2 \left(c_2 c_0^2 m^2+k\right)+6 c_3 c_0^3 m^2 r_h+c_1 c_0 m^2 r_h^3+6 c_4 c_0^4 m^2+4 \pi  P r_h^4\right]}{4 \pi }-\frac{8 \pi  Q^2}{V_4 r_h^4}-\frac{V_4 r_h^3}{\tau }.
\label{eq3.2.3}
\end{eqnarray}

\noindent Zero points are determined by the equation $\phi^{r_h} = 0$. Then the relation between $r_h$ and $\tau$ is 

\begin{eqnarray}
\tau =\frac{4 \pi  r_h^7}{-32 \pi ^2 Q^2/V_4^2 +6 c_4 c_0^4 m^2 r_h^4 + 6  c_3 c_0^3 m^2 r_h^5+ 3 \left(c_2 c_0^2 m^2+k\right) r_h^6+ c_1 c_0 m^2 r_h^7 +4 \pi  P r_h^8}.
\label{eq3.2.4}
\end{eqnarray}

\begin{figure}[h]
	\centering
	\begin{minipage}[t]{0.27\textwidth}
		\centering
		\includegraphics[scale=0.27]{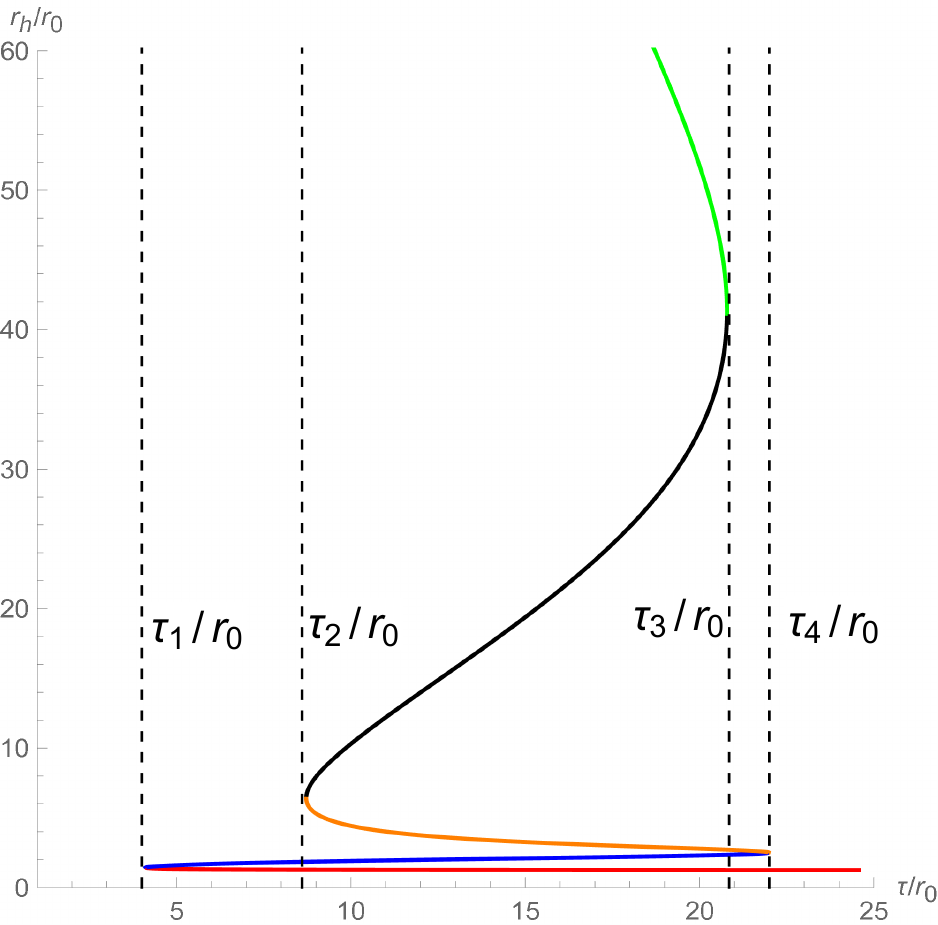}
	\end{minipage}
	\begin{minipage}[t]{0.27\textwidth}
		\centering
		\includegraphics[scale=0.27]{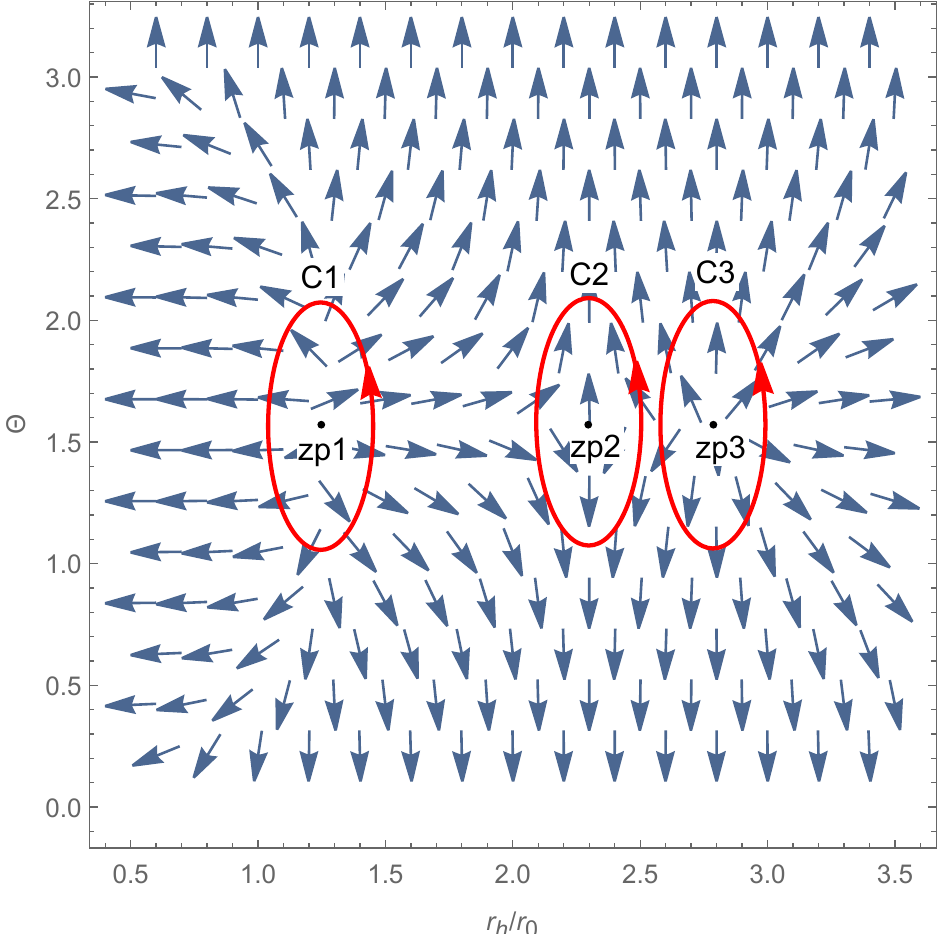}
	\end{minipage}
	\begin{minipage}[t]{0.27\textwidth}
		\centering
		\includegraphics[scale=0.27]{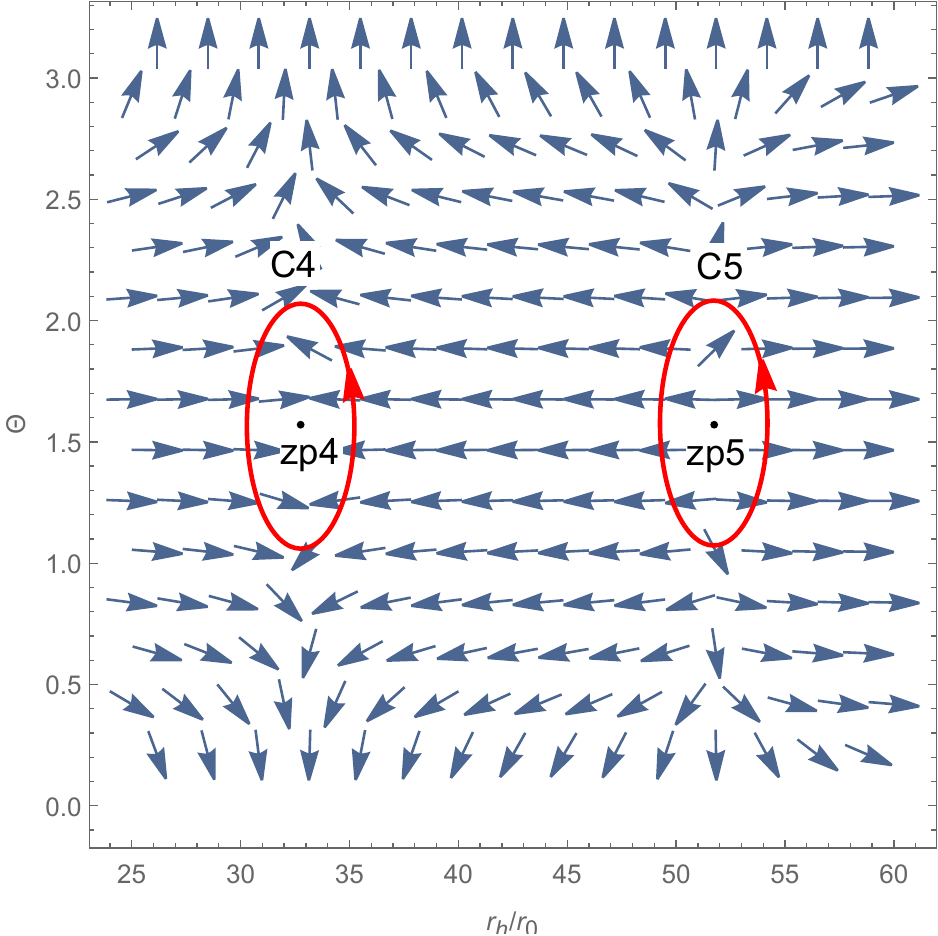}
	\end{minipage}
\caption{Topological properties of the six-dimensional charged black hole, where $a_6=\frac{1}{32\pi ^2}$, $b_6=25$, $c_6=-20$, $d_6=8$, $e_6 =-0.5$, $k=1$ and $Pr_0^{2}=0.001$.  Zero points of the vector $\phi^{r_h}$ in the plane $r_h - \tau$ are plotted in the left picture. The unit vector field $n$ on a portion of the plane $r_h - \Theta$ at $\tau /r_0=20 $ is plotted in the middle and right pictures. Zero points are at $(r_h/r_0 ,\Theta)$=($1.25, \pi/2$), ($2.29, \pi/2$), ($2.79, \pi/2$), ($32.76, \pi/2$) and ($51.75, \pi/2$), respectively.}	
	\label{fig:4.1}
\end{figure}

\begin{figure}[h]
	\centering
	\begin{minipage}[t]{0.48\textwidth}
		\centering
		\includegraphics[scale=0.3]{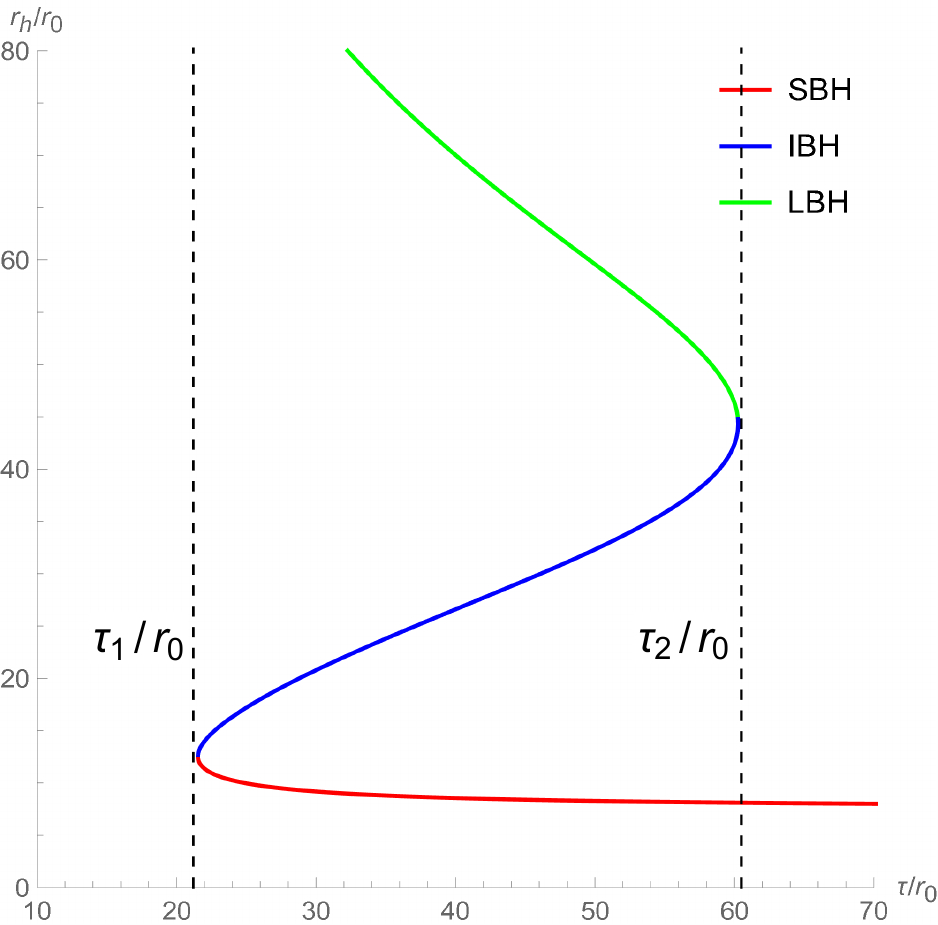}
	\end{minipage}
	\begin{minipage}[t]{0.48\textwidth}
		\centering
		\includegraphics[scale=0.3]{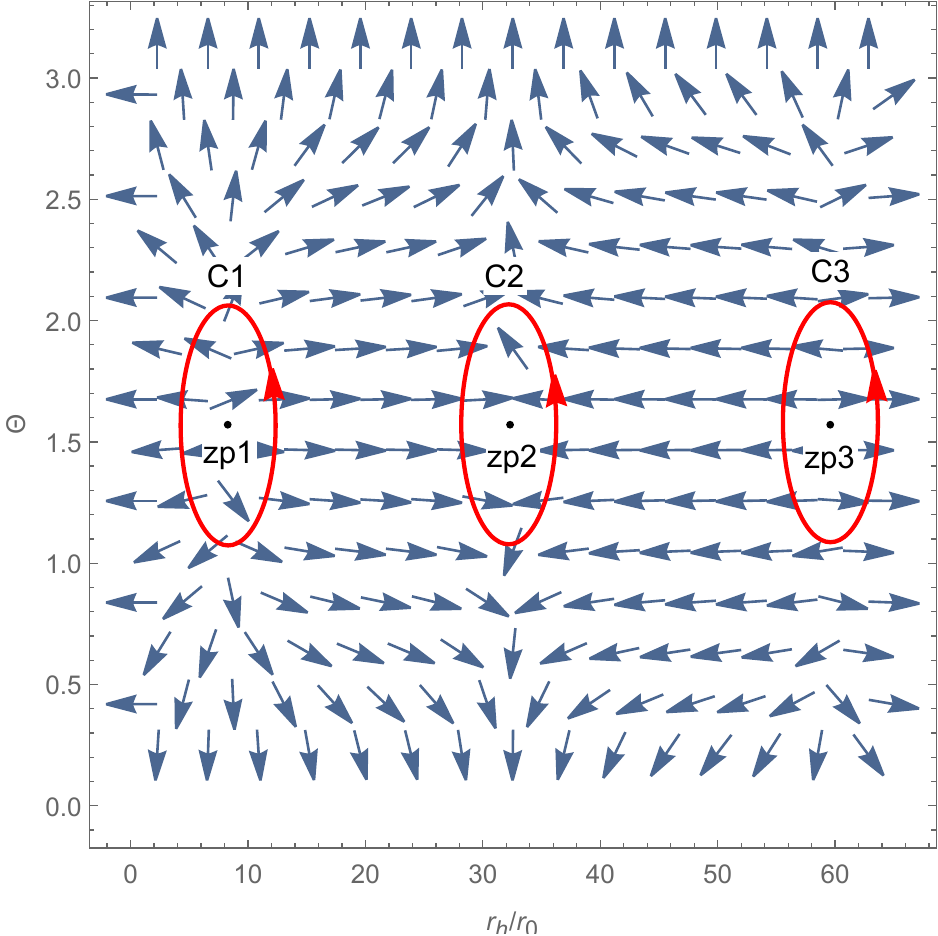}
	\end{minipage}
\caption{Topological properties of the six-dimensional charged black hole, where $a_6=\frac{1}{32\pi ^2}$, $b_6=-20$, $c_6=-30$, $d_6=10$, $e_6 =-1$, $k=1$ and $Pr_0^{2}=0.001$. Zero points of the vector $\phi^{r_h}$ in the plane $r_h - \tau$ are plotted in the left picture. The unit vector field $n$ on a portion of the plane $r_h - \Theta$ at $\tau /r_0=20$ is plotted in the right picture. Zero points are at $(r_h/r_0 ,\Theta)$=($8.27, \pi/2$), ($32.32, \pi/2$) and ($59.58, \pi/2$), respectively.}	
	\label{fig:4.2}
\end{figure}

\begin{figure}[h]
	\centering
	\begin{minipage}[t]{0.48\textwidth}
		\centering
		\includegraphics[scale=0.3]{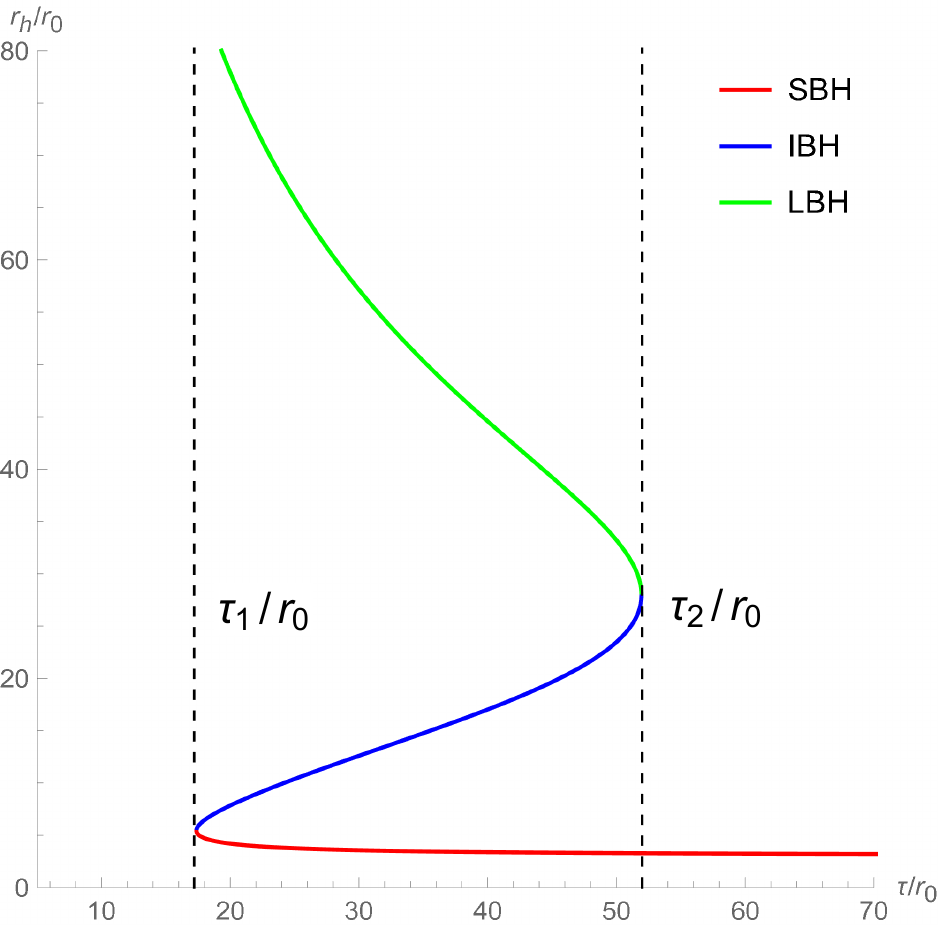}
	\end{minipage}
	\begin{minipage}[t]{0.48\textwidth}
		\centering
		\includegraphics[scale=0.3]{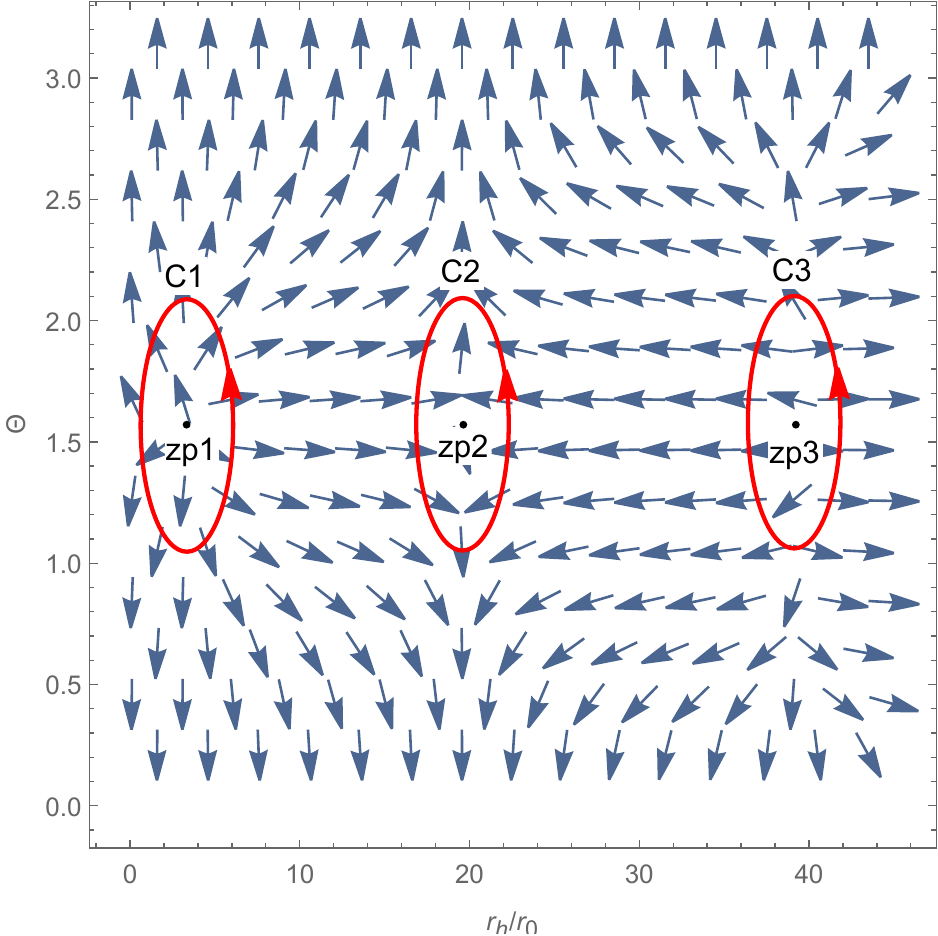}
	\end{minipage}
\caption{Topological properties of the six-dimensional uncharged black hole, where $a_6=0$,  $b_6=-1$, $c_6=-5$, $d_6= 3$, $e_6 = -0.5$, $k=1$ and $Pr_0^{2}=0.001$. Zero points of the vector $\phi^{r_h}$ in the plane $r_h - \tau$ are plotted in the left picture. The unit vector field $n$ on a portion of the plane $r_h - \Theta$ at $\tau /r_0=45 $ is plotted in the right picture. Zero points are at $(r_h/r_0 ,\Theta)$=($3.33, \pi/2$), ($19.64, \pi/2$) and ($39.23, \pi/2$), respectively.}	
	\label{fig:4.3}
\end{figure}

\begin{figure}[h]
	\centering
	\begin{minipage}[t]{0.27\textwidth}
		\centering
		\includegraphics[scale=0.27]{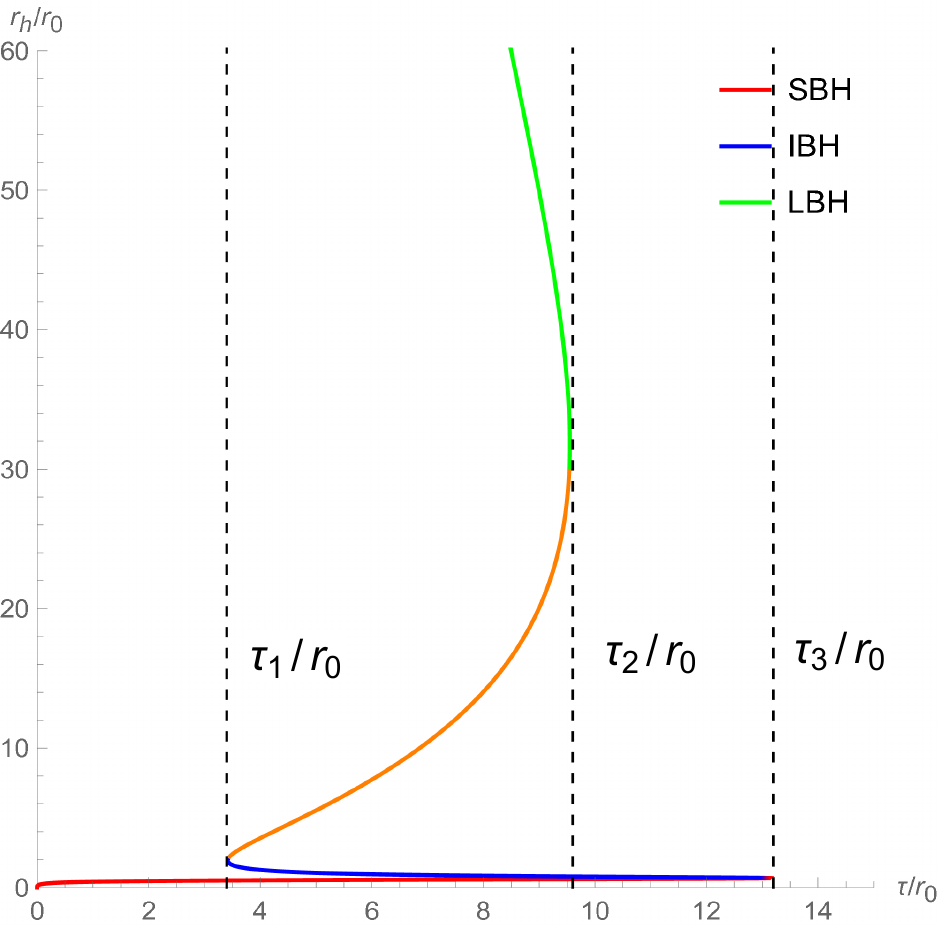}
	\end{minipage}
	\begin{minipage}[t]{0.27\textwidth}
		\centering
		\includegraphics[scale=0.27]{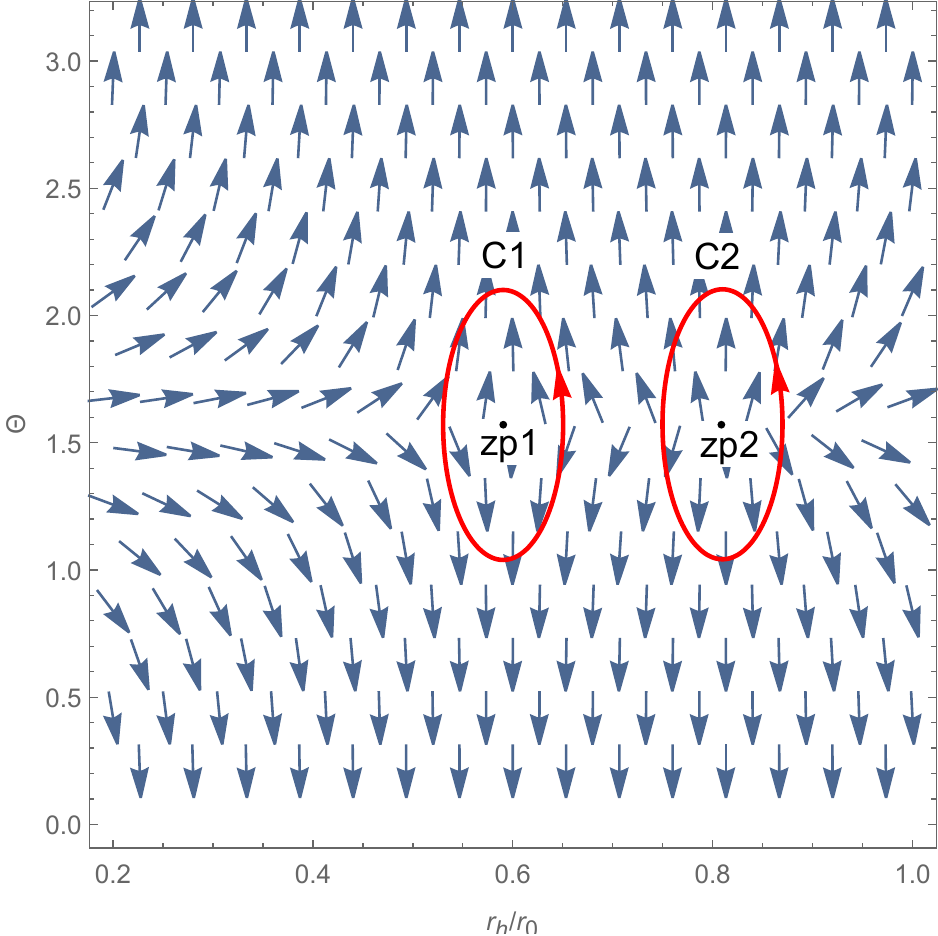}
	\end{minipage}
	\begin{minipage}[t]{0.27\textwidth}
		\centering
		\includegraphics[scale=0.27]{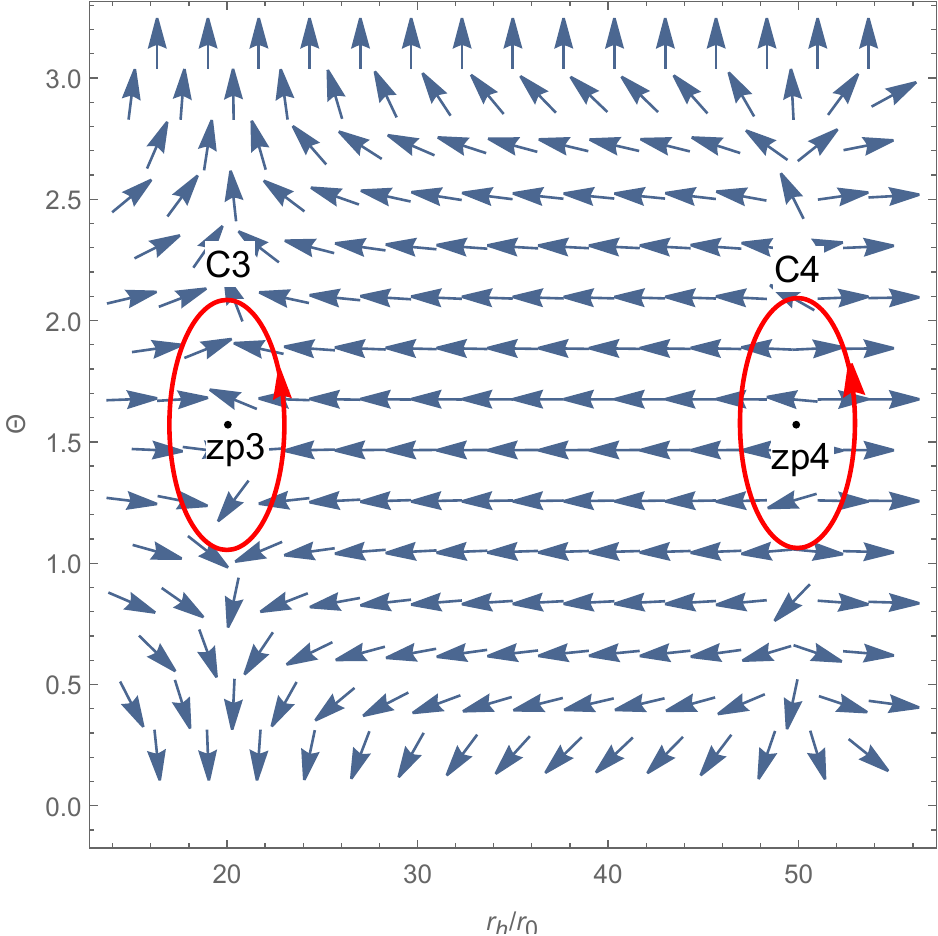}
	\end{minipage}
\caption{Topological properties of the six-dimensional uncharged black hole, where $a_6=0$,  $b_6=-1$, $c_6= -3$, $d_6= 3.6$, $e_6 = 0.5$, $k=1$, $Pr_0^{2}=0.001$.  Zero points of the vector $\phi^{r_h}$ in the plane $r_h - \tau$ are plotted in the left picture. The unit vector field $n$ on a portion of the plane $r_h - \Theta$ at $\tau /r_0=20 $ is plotted in the middle and right pictures. Zero points are at $(r_h/r_0 ,\Theta)$=($0.59, \pi/2$), ($0.81, \pi/2$), ($20.05, \pi/2$) and ($49.88, \pi/2$), respectively.}	
	\label{fig:4.4}
\end{figure}

\noindent For convenience, we let $a_6=\frac{Q^2}{V_3^2r_0^{6}}$, $b_6=c_0^4 c_4 m^2/r_0^2$,  $ c_6=c_0^3 c_3 m^2/r_0$, $d_6= c_0^2 c_2 m^2$, $e_6=c_0 c_1 m^2 r_0$,  $k=1$ and $Pr_0^{2}=0.001$ and draw Figure \ref{fig:4.1} - Figure \ref{fig:4.4} in this subsection.

We first study the topological properties of the six-dimensional charged black hole. In Figure \ref{fig:4.1}, there are two generation points and two annihilation points. The generation points are at $\tau_1/r_0$ and $\tau_2/r_0$, respectively. The annihilation points are at $\tau_1/r_0$ and $\tau_2/r_0$, respectively. These points divide the black hole into five branches, namely, three stable black hole branches and two unstable black hole branches, which leads to the topological number as $ 1$. Changing the parameters' values, we find that there are three black hole branches and three zero points in Figure \ref{fig:4.2}. It is easily to get that the number is $ 1$. Although their numbers are the same in Figure \ref{fig:4.1} and \ref{fig:4.2}, they have different phase transitions. Clearly, the number for this black hole is $1$.

For the six-dimensional uncharged black hole, there are three black hole branches and three zero points in Figure \ref{fig:4.3}, and then its topological number is $1$. This black hole has a LBH/SBH phase transition. In Figure \ref{fig:4.4}, there are one generation point and two annihilation points, which divide the black hole into two stable black hole branches and two unstable black hole branches. Four zero points are at $(r_h/r_0 ,\Theta)$=($0.59, \pi/2$), ($0.81, \pi/2$), ($20.05, \pi/2$) and ($49.88, \pi/2$), respectively. Thus its topological number is $=0$. Comparing Figure  \ref{fig:4.3} and \ref{fig:4.4}, we find that the number for the uncharged black hole is $0$ or $1$. This result is different from that for the six-dimensional charged black hole.

\subsection{Topological numbers for seven-dimensional black holes}

We continue to study the topology for the seven-dimensional black holes. Then $n=5$ in the metric (\ref{eq2.2}). The mass and entropy are given by

\begin{eqnarray}
	M=\frac{5 V_5 r_h^4}{16 \pi }\left[\frac{12 c_4 c_0^4 m^2}{r_h^2}+\frac{4 c_3 c_0^3 m^2}{r_h}+\frac{1}{5} c_0 c_1 m^2 r_h+c_2 c_0^2 m^2+\frac{8 \pi  Pr_h^2}{15}+\frac{(16 \pi  Q)^2}{40 V_5^2 r_h^8}+k\right],
\label{eq3.3.1}
\end{eqnarray}

\begin{eqnarray}
S=\frac{1}{4} V_5 r_h^5
\label{eq3.3.2}
\end{eqnarray}

\noindent respectively. We use Eqs. (\ref{eq3.1}), (\ref{eq3.3.1}) and (\ref{eq3.3.2}), and obtain the generalized off-shell free energy

\begin{eqnarray}
\mathcal{F}=\frac{5 V_5 r_h^4}{16 \pi }\left[\frac{12 c_4 c_0^4 m^2}{r_h^2}+\frac{4 c_3 c_0^3 m^2}{r_h}+\frac{c_0 c_1 m^2r_h}{5}+c_2 c_0^2 m^2+k+\frac{8 \pi  P r_h^2}{15} +\frac{(8 \pi  Q)^2}{10 V_5^2 r_h^8}\right]-\frac{V_5 r_h^5}{4 \tau }.
\label{eq3.3.3}
\end{eqnarray}

\noindent Adopting the definition $\phi^{r_h}=\frac{\partial \mathcal{F}}{\partial r_h}$, we get the component of the vector, which is

\begin{eqnarray}
\phi^{r_h} &= & \frac{5 V_5 r_h}{16 \pi }\left[(c_2 c_0^2 m^2+k)r_h^2 +c_0 c_1 m^2  r_h^3 + 60 c_0^3 c_3 m^2 r_h  + 24 c_0^4 c_4 m^2 \right] +P V_5 r_h^5 \nonumber\\
&& -\frac{8 \pi  Q^2}{V_5 r_h^5}-\frac{5 V_5 r_h^4}{4 \tau }.
\label{eq3.3.4}
\end{eqnarray}

\noindent We solve $\phi^{r_h} = 0$ and obtain the relation between $r_h$ and $\tau$,

\begin{eqnarray}
\tau = \frac{20 \pi  V_5^2 r_h^9}{16 \pi  P V_5^2 r_h^{10}+ 5 c_0 c_1 m^2 V_5^2 r_h^9+20 (c_0^2 c_2 m^2 +k)V_5^2 r_h^8+60 c_0^3 c_3 m^2 V_5^2 r_h^7+120 c_0^4 c_4 m^2 V_5^2 r_h^6 -128 \pi ^2 Q^2} =
\label{eq3.3.5}
\end{eqnarray}

\begin{figure}[h]
	\centering
	\begin{minipage}[t]{0.29\textwidth}
		\centering
		\includegraphics[scale=0.3]{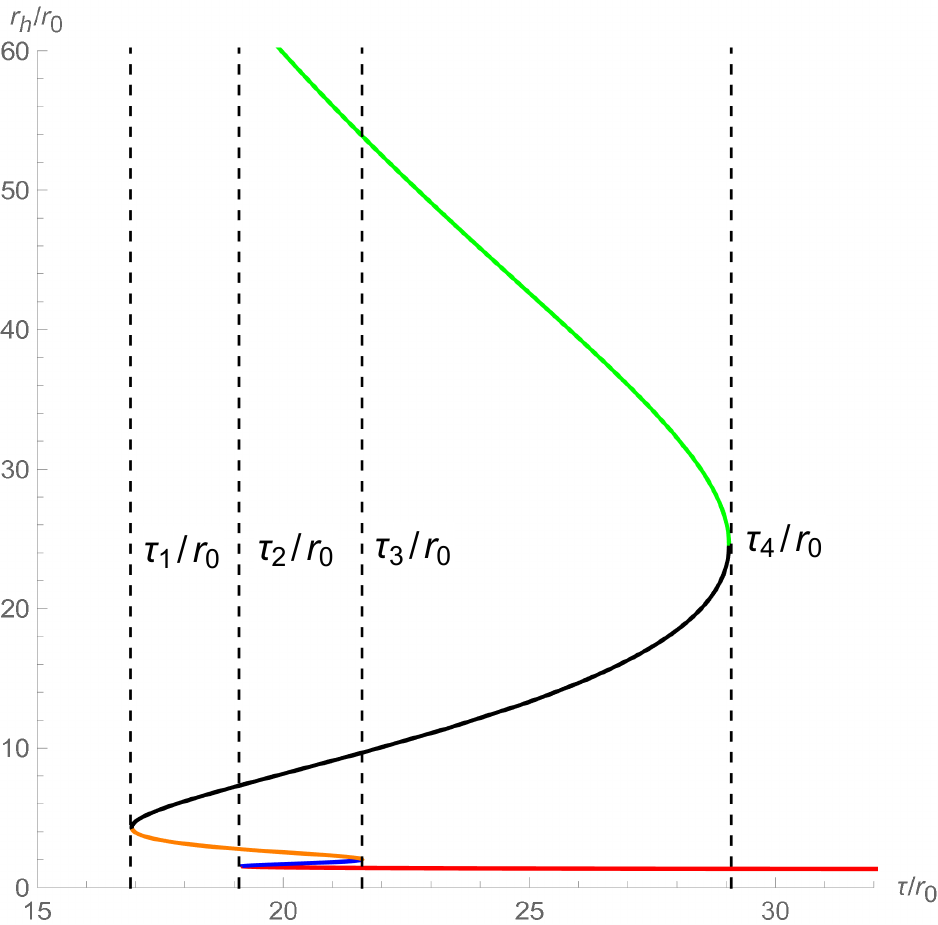}
	\end{minipage}
	\begin{minipage}[t]{0.29\textwidth}
		\centering
		\includegraphics[scale=0.3]{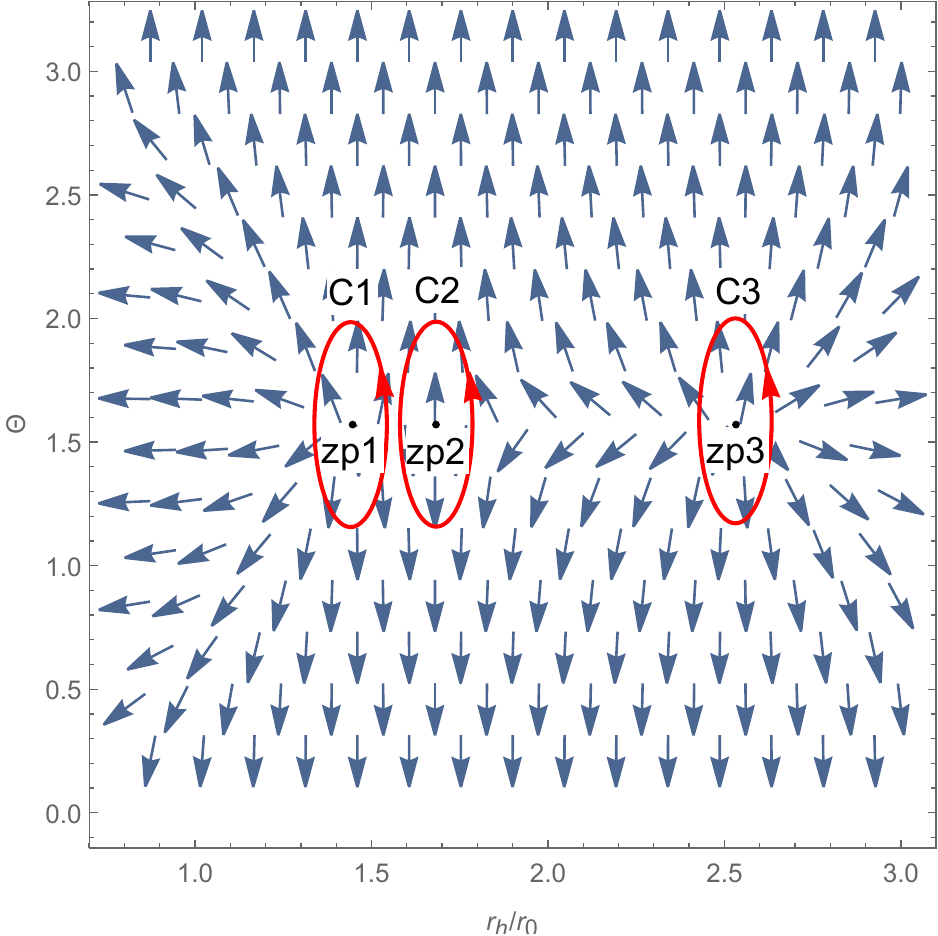}
	\end{minipage}
	\begin{minipage}[t]{0.29\textwidth}
		\centering
		\includegraphics[scale=0.3]{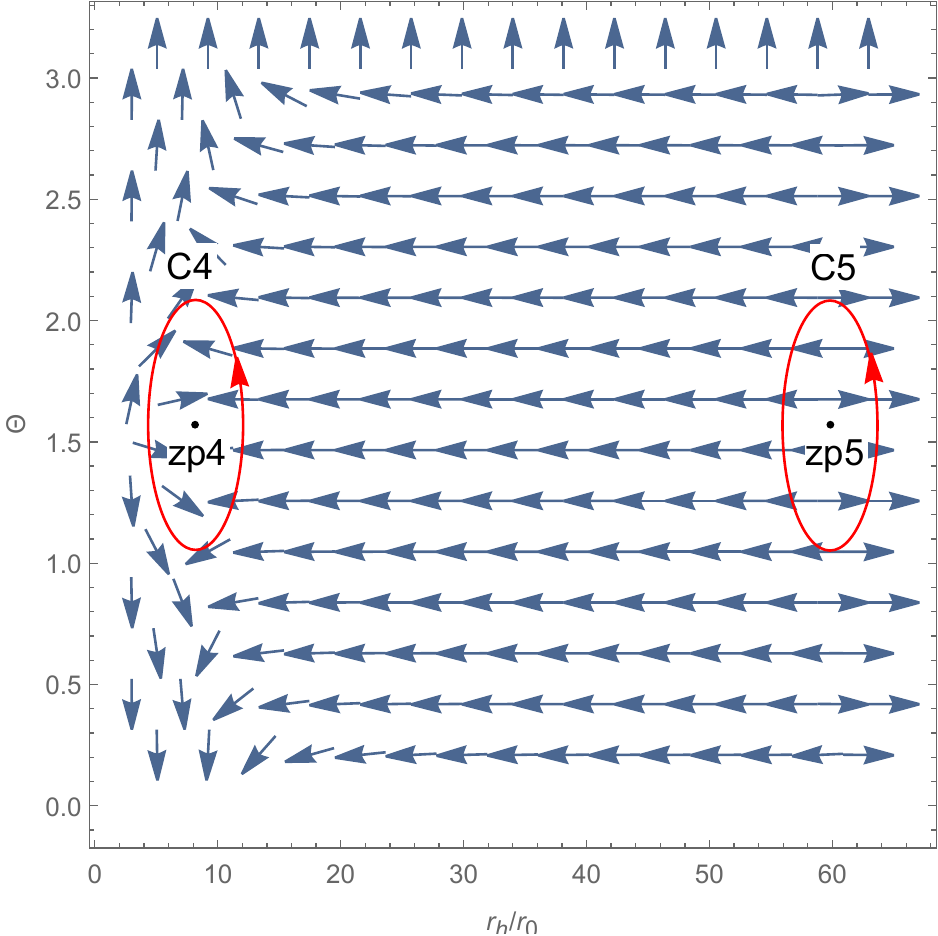}
	\end{minipage}
\caption{Topological properties of the seven-dimensional charged black hole, where $a_7= \frac{1}{128\pi ^2}$, $b_7 =1$, $c_7= -2.1$, $d_7= 1$, $e_7=-0.1$, $k=1$ and $Pr_0^{2}=0.001$.  Zero points of the vector $\phi^{r_h}$ in the plane $r_h - \tau$ are plotted in the left picture. The unit vector field $n$ on a portion of the plane $r_h - \Theta$ at $\tau /r_0=20 $ is plotted in the middle and right pictures. Zero points are at $(r_h/r_0 ,\Theta)$=($1.45 , \pi/2$), ($1.68, \pi/2$), ($2.53, \pi/2$), ($8.15, \pi/2$) and ($59.84, \pi/2$), respectively.}	
	\label{fig:5.1}
\end{figure}

\begin{figure}[h]
	\centering
	\begin{minipage}[t]{0.48\textwidth}
		\centering
		\includegraphics[scale=0.3]{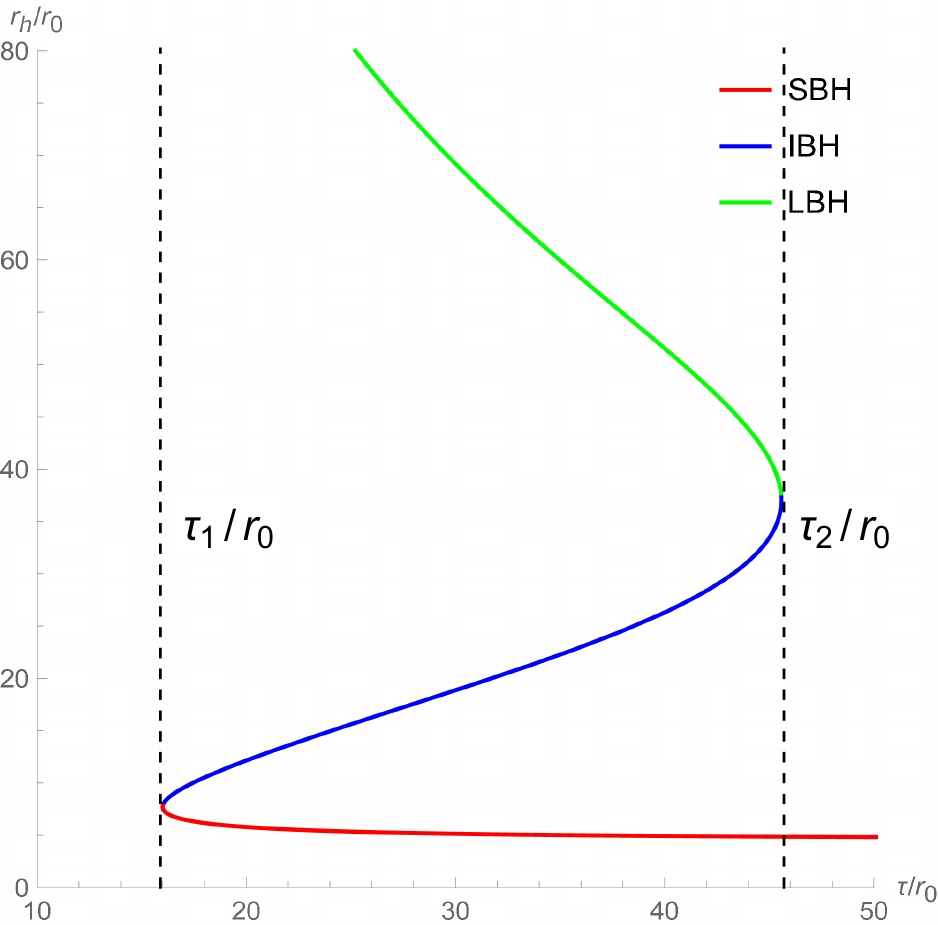}
	\end{minipage}
	\begin{minipage}[t]{0.48\textwidth}
		\centering
		\includegraphics[scale=0.3]{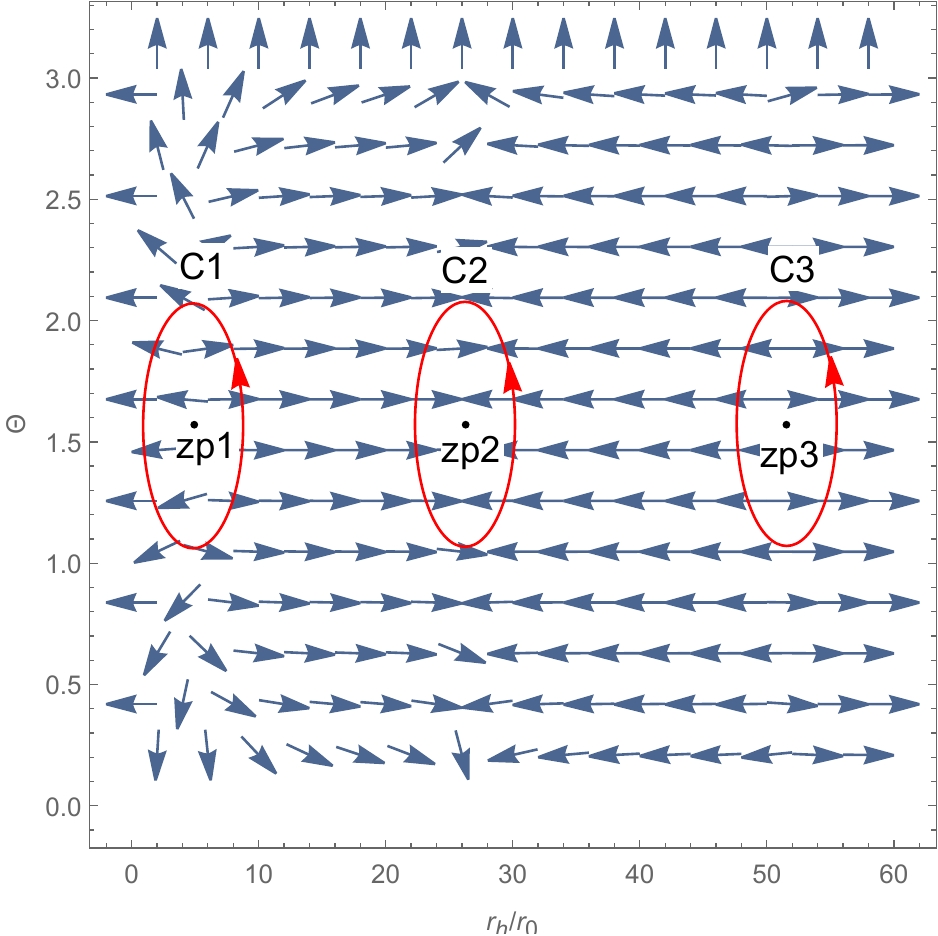}
	\end{minipage}
\caption{Topological properties of the seven-dimensional charged black hole, where  $a_7= \frac{1}{128\pi ^2}$, $b_7 = -5$, $c_7= -3$, $d_7= 3$, $e_7=-0.5$, $k=1$ and $Pr_0^{2}=0.001$. Zero points of the vector $\phi^{r_h}$ in the plane $r_h - \tau$ are plotted in the left picture. The unit vector field $n$ on a portion of the plane $r_h - \Theta$ at $\tau /r_0=40 $ is plotted in the right picture. Zero points are at $(r_h/r_0 ,\Theta)$=($4.93, \pi/2$), ($26.30, \pi/2$) and ($51.54, \pi/2$), respectively.}	
	\label{fig:5.2}
\end{figure}

\begin{figure}[h]
	\centering
	\begin{minipage}[t]{0.29\textwidth}
		\centering
		\includegraphics[scale=0.3]{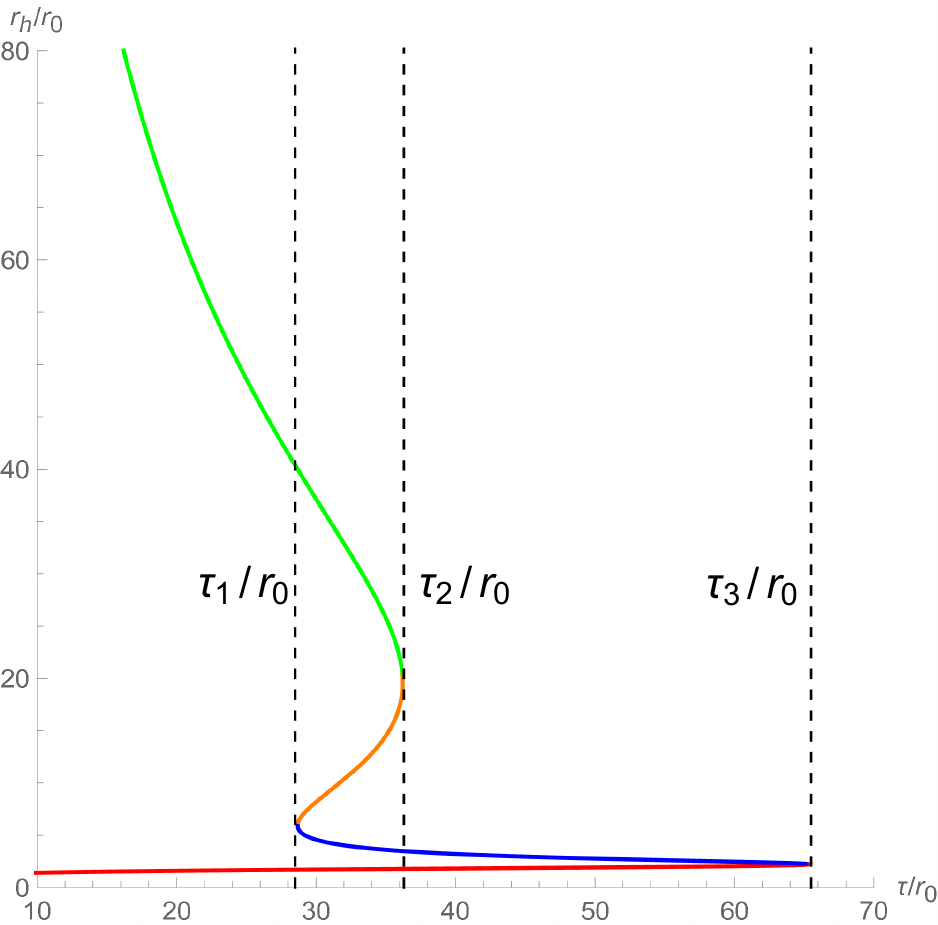}
	\end{minipage}
	\begin{minipage}[t]{0.29\textwidth}
		\centering
		\includegraphics[scale=0.3]{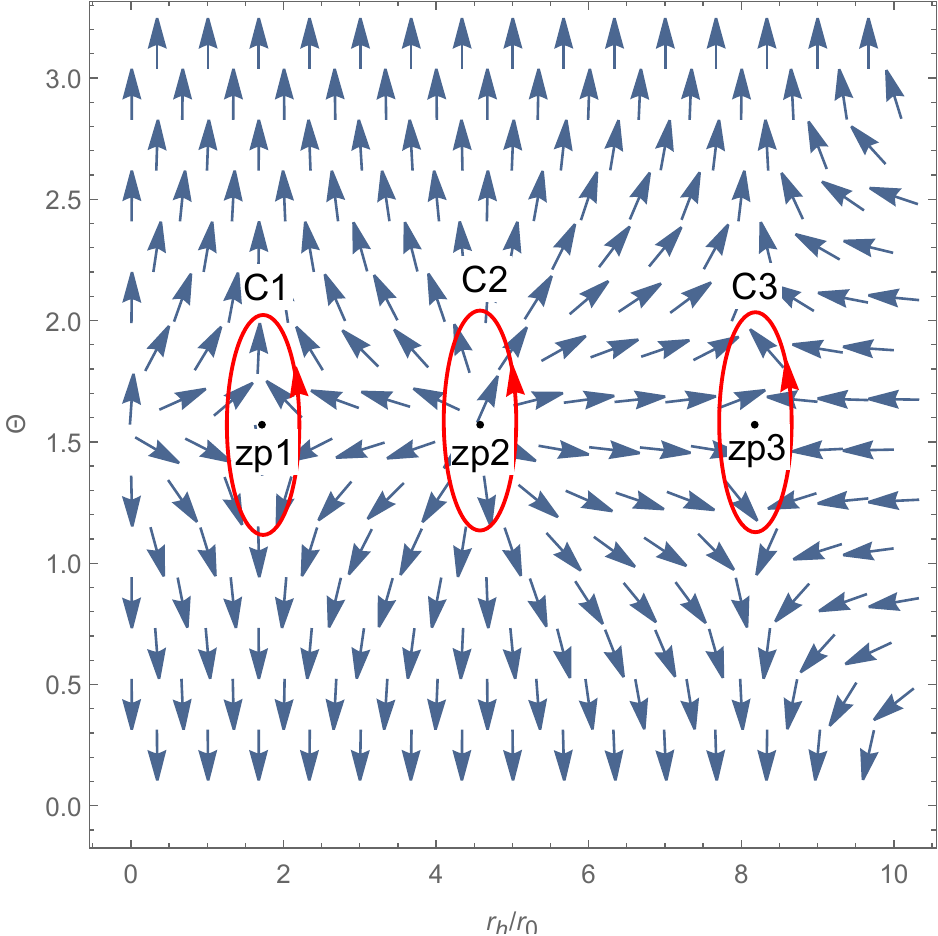}
	\end{minipage}
	\begin{minipage}[t]{0.29\textwidth}
		\centering
		\includegraphics[scale=0.3]{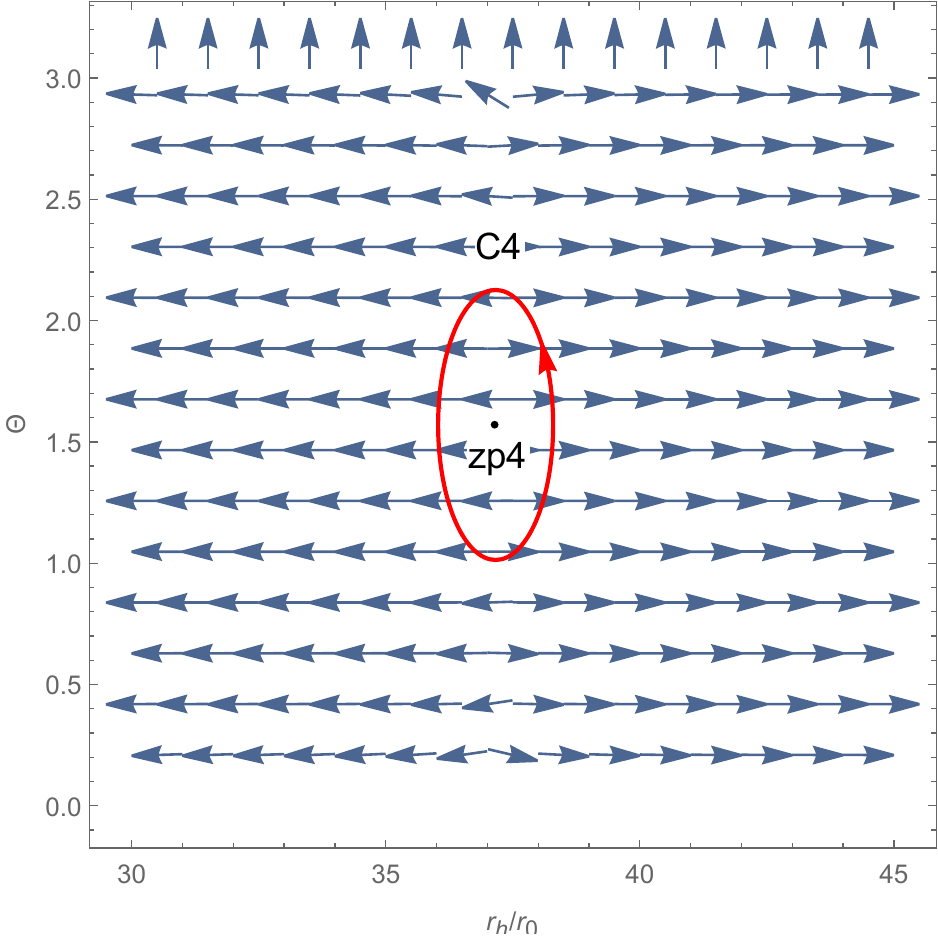}
	\end{minipage}
\caption{Topological properties of the seven-dimensional uncharged black hole, where  $a_7= 0$, $b_7 = 1$, $c_7= -1.9$, $d_7= 0.5$, $e_7=-0.1$, $k=1$ and $Pr_0^{2}=0.001$. Zero points of the vector $\phi^{r_h}$ in the plane $r_h - \tau$ are plotted in the left picture. The unit vector field $n$ on a portion of the plane $r_h - \Theta$ at $\tau /r_0=40 $ is plotted in the middle and right pictures. Zero points are at $(r_h/r_0 ,\Theta)$=($1.72, \pi/2$), ($4.58, \pi/2$), ($8.18, \pi/2$) and ($37.14, \pi/2$), respectively.}	
	\label{fig:5.3}
\end{figure}

\begin{figure}[h]
	\centering
	\begin{minipage}[t]{0.48\textwidth}
		\centering
		\includegraphics[scale=0.3]{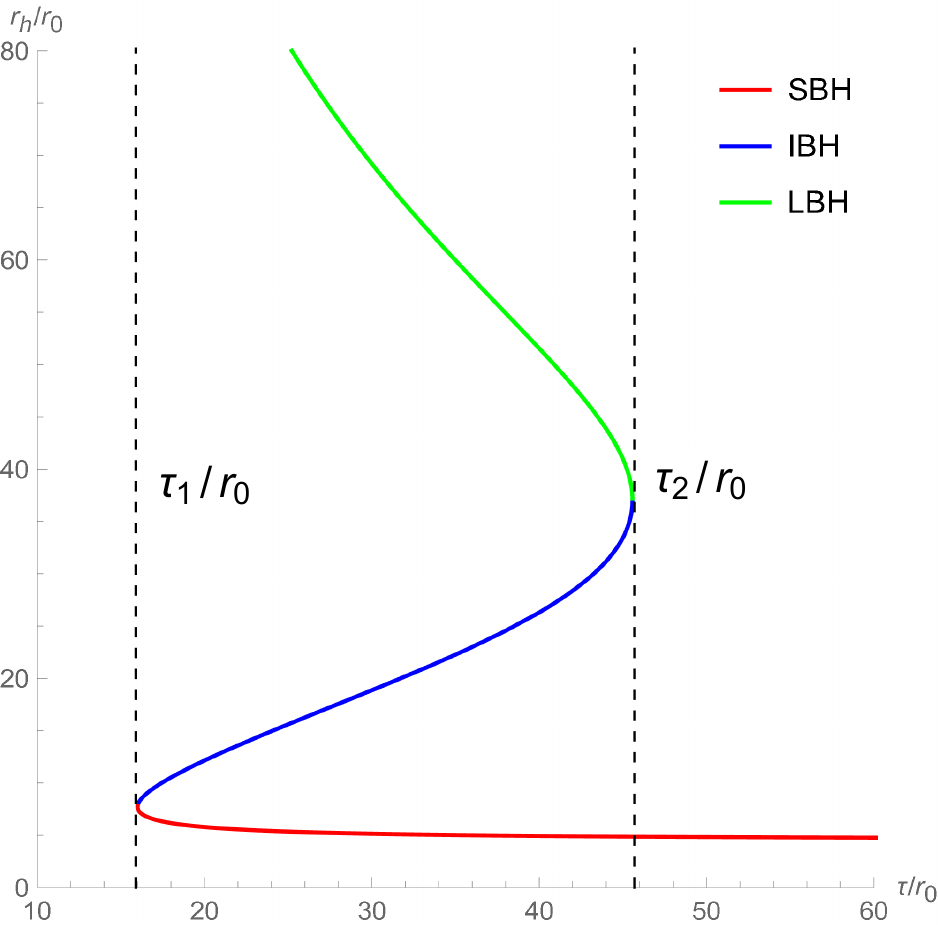}
	\end{minipage}
	\begin{minipage}[t]{0.48\textwidth}
		\centering
		\includegraphics[scale=0.3]{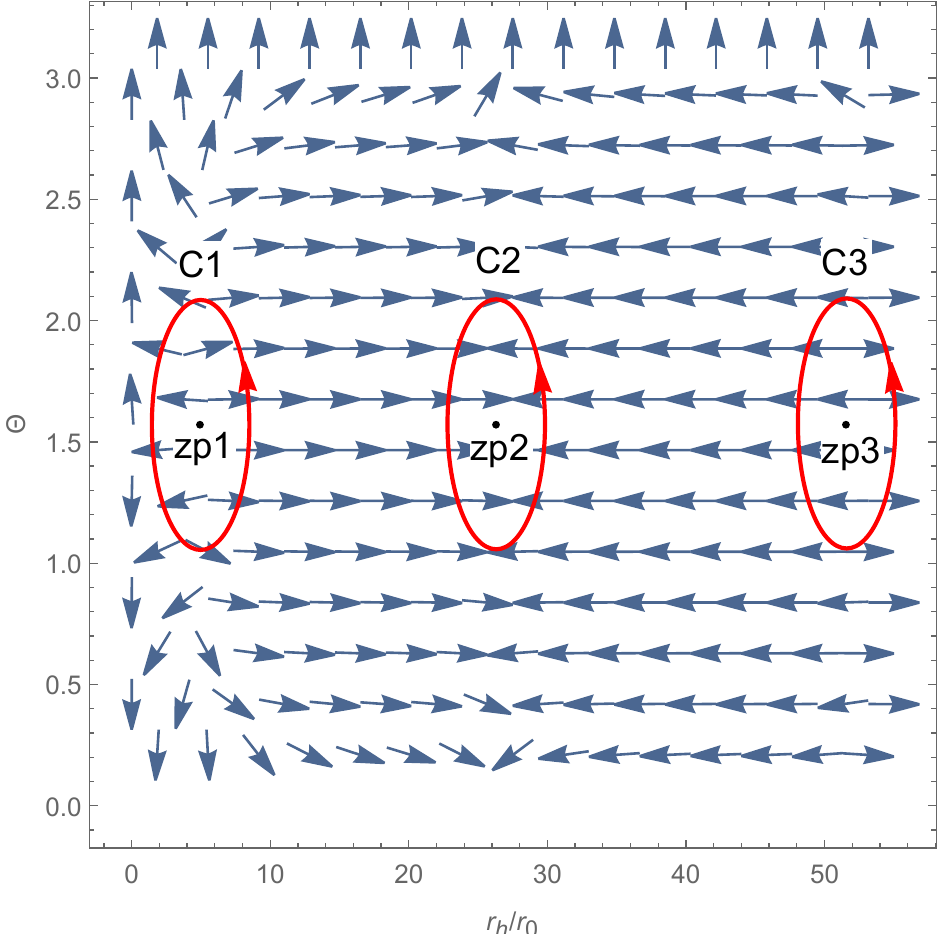}
	\end{minipage}
\caption{Topological properties of the seven-dimensional uncharged black hole, where $a_7= 0$, $b_7 = -5$, $c_7= -3$, $d_7= 3$, $e_7=-0.5$, $k=1$ and $Pr_0^{2}=0.001$. Zero points of the vector $\phi^{r_h}$ in the plane $r_h - \tau$ are plotted in the left picture. The unit vector field $n$ on a portion of the plane $r_h - \Theta$ at $\tau /r_0=40 $ is plotted in the right picture. Zero points are at $(r_h/r_0 ,\Theta)$=($4.93, \pi/2$), ($26.29, \pi/2$) and ($51.54, \pi/2$), respectively.}	
	\label{fig:5.4}
\end{figure}

\noindent We order $a_7=\frac{Q^2}{V_5^2r_0^{8}}$, $b_7=c_0^4 c_4 m^2/r_0^2$,  $ c_7=c_0^3 c_3 m^2/r_0$, $d_7= c_0^2 c_2 m^2$, $e_6=c_0 c_1 m^2 r_0$,  $k=1$ and $Pr_0^{2}=0.001$ and plot Figure \ref{fig:5.1} - Figure \ref{fig:5.4}.

In Figure \ref{fig:5.1}, there are two generation points ($\tau_1/r_0$  and $\tau_2/r_0$) and two annihilation points ($\tau_1/r_0$  and $\tau_2/r_0$). They divide the black hole into three stable black hole branches and two unstable black hole branches. Its topological number is $1$. When the temperature is low, the black hole is large. When the temperature is high, the black hole is small. There are only one generation point and one annihilation point in Figure \ref{fig:5.2}. These points divide the black hole into three branches and three zero points appear. Therefore, the number is $1$. Although this black hole has different branches and numbers of zero points for the different parameters' values, it has the same number.

When the black hole is uncharged, its topology is discussed in Figure \ref{fig:5.3} and Figure \ref{fig:5.4}. In the left picture of Figure \ref{fig:5.3}, two annihilation points and one generation point divide the black hole into two stable black hole branches and two unstable black hole branches, which yields the number is $0$. In Figure \ref{fig:5.4}, there are three branches and zero points. Its topological number is $1$. These two figures shows that the topological number for this black hole is $0$ or $1$.

\section{Conclusion and discussion}\label{Sec4}

In this work, we studied the topological numbers of the five-, six- and seven-dimensional anti-de Sitter black holes in the ghost-free massive gravity. The numbers for these black holes obtained in the work are listed in Table 1. In the study, we first used the different values of the five-dimensional black holes' parameters and calculated their topological numbers. Due to the different values, the numbers of generation points, annihilation points and zero points are different. However, these data lead to the same number for the five-dimensional charged black hole, while the number for the uncharged black holes  is $0$ or $1$. Its specific values are determined by the values of the black holes' parameters. This situation also occurs in the six- and seven-dimensional black holes.

\begin{table}[H]
   \centering
   \begin{tabular}{|c|c|c|}
   	\hline
   	Black hole solutions &Topological numbers \\
   	\hline
   	5d charged BH & 1 \\
   	\hline
   	5d uncharged BH & $0$ or $1$\\
   	\hline
   	6d charged BH& 1\\
   	\hline
   	6d uncharged BH&$0$ or $1$\\
   	\hline
   	7d charged BH& 1 \\
   	\hline
   	7d uncharged BH& $0$ or $1$\\
   	\hline
   \end{tabular}
   \caption{5d/6d/7d represent ``a five-dimenional/six-dimenional/seven-dimensional'',  BH is the abbreviation for the black hole.}
   \label{t.1.4}
\end{table}

In the work, we only studied the topological numbers of the spherically symmetric (i.e. $k=1 $) black holes. The reason is that $k$ and $c_ 0^2c_2 m^2 $ appear together in the generalized free energy in the form of $k +c_ 0^2c_2 m^2 $, where $k$ characterizes the horizon curvature and $c_2 m^2 $ is the coefficient of the second term of massive potential associated with the graviton mass. In the calculation, we considered the combined effect of these two physical quantities. That is to say, as long as the sum of the two quantities' values is a constant, the topological numbers calculated in this paper are the same. For example, $k = 1$ and $c_ 0^2c_2 m^2 = 3$ in Figure \ref{fig:5.4}, which yields $k + c_ 0^2c_2 m^2  = 4$. If we order $k = -1$ or $0$, thus $c_ 0^2c_2 m^2 = 5$ or $4$, and the same result can be gotten in Figure \ref{fig:5.4}. In Figure \ref{fig:5.3}, $k = 1$ and $c_ 0^2c_2 m^2 = 0.5$, which leads to $k + c_ 0^2c_2 m^2  = 1.5$. When $k = -1$ or $0$, and then we get $c_ 0^2c_2 m^2 = 2.5$ or $1.5$, and the same result in Figure \ref{fig:5.3}. Therefore, the numbers for the Ricci flat and hyperbolic, charged and uncharged black holes in the five-, six- and seven-dimensional massive gravity are also obtained. The charged black holes have a same number. For the uncharged black holes, the numbers are $0$ or $1$. The number for the four-dimensional black hole in the dRGT massive gravity was calculated in \cite{TS}. He found that the number for the uncharged black hole is $=0$.

In \cite{GP1}, Gogoi and Phukon have studied the topological properties of the four-dimensional Dyonic AdS black hole in the different ensembles and found that the topological class of this black hole is ensemble dependent. Therefore, it is quite meaningful to study the topological classes of high-dimensional black holes in the massive gravity in different ensembles, which may lead to an interesting result.

\end{document}